\documentclass[manuscript,nonacm]{acmart}
\usepackage{subcaption}
\usepackage{soul}
\usepackage{multirow}
\usepackage{tcolorbox}
\usepackage{enumitem}
\usepackage{rotating}
\usepackage{xcolor,colortbl}
\definecolor{lavender}{rgb}{0.9, 0.9, 0.98}

\begin{document}

\title[Understanding the Characteristics and Role of Argumentation in Open-Source Software Usability Discussions]{What Pulls the Strings? Understanding the Characteristics and Role of Argumentation in Open-Source Software Usability Discussions}


\author{Arghavan Sanei}
\email{arghavan.sanei@polymtl.ca}
\orcid{0009-0007-9601-8094}
\affiliation{%
  \institution{Polytechnique Montreal}
  \city{Montreal}
  \state{Quebec}
  \country{Canada}
}

\author{Chaima Amiri}
\email{chaima.amiri@polymtl.ca}
\affiliation{%
  \institution{Polytechnique Montreal}
  \city{Montreal}
  \state{Quebec}
  \country{Canada}
}

\author{Atefeh Shokrizadeh}
\email{atefeh.shokrizadeh@polymtl.ca}
\orcid{0009-0004-0743-6875}
\affiliation{%
  \institution{Polytechnique Montreal}
  \city{Montreal}
  \state{Quebec}
  \country{Canada}
}

\author{Jinghui Cheng}
\email{Jinghui.cheng@polymtl.ca}
\orcid{0000-0002-8474-5290}
\affiliation{%
  \institution{Polytechnique Montreal}
  \city{Montreal}
  \state{Quebec}
  \country{Canada}
}

\renewcommand{\shortauthors}{Sanei et al.}

\begin{abstract}
\textbf{Abstract}: 
The usability of open-source software (OSS) is important but frequently overlooked in favor of technical and functional complexity. Argumentation can be a pivotal device for diverse stakeholders in OSS usability discussions to express opinions and persuade others. However, the characteristics of argument discourse in those discussions remain unknown, resulting in difficulties in providing effective support for discussion participants. We address this through a comprehensive analysis of argument discourse and quality in five OSS projects. Our results indicated that usability discussions are predominantly argument-driven, although their qualities vary. Issue comments exhibit lower-quality arguments than the issue posts, suggesting a shortage of collective intelligence about usability in OSS communities. Moreover, argument discourse and quality have various impacts on the subsequent behavior of participants. Overall, this research offers insights to help OSS stakeholders build more effective arguments and eventually improve OSS usability. These insights can also inform studies about other distributed collaborative communities.
\end{abstract}


\maketitle

\section{Introduction}
\label{sec:intro}
Usability of open source software (OSS) projects often receive less attention compared to other aspects, such as security, functionalities, and performance, leading to overall inferior usability in OSS applications~\cite{wang2020argulens, Hellman2021Facilitating, nichols2006usability,rajanen2015power}. The majority of the OSS communities are mostly focusing on the technical and functional problems of their projects while putting the user interface (UI) and user experience (UX) design along with the associated usability concerns a lower priority~\cite{wang2020argulens, Wang2022IEEESoftware}. These factors rendered usability a long-lasting and challenging problem for many OSS projects~\cite{nichols2006usability, Andreasen2006, raza2012users}. One important place of raising, discussing, and resolving these types of problems in OSS is through issue tracking systems (ITSs)~\cite{Wang2022IEEESoftware, sanei2023characterizing, cheng2018open}, such as GitHub Issues~\cite{githubGitHubIssues}. Usability-related discussions in ITSs often involve diverse participants with a wide spectrum of backgrounds and experiences (e.g., developers, maintainers, designers, and end users~\cite{Cheng2019CHASE, cheng2018open}).

With the diverse participants and the intricacies of usability problems in OSS, people who raise and discuss usability-related topics on ITSs (or ``usability issues'' in short) frequently need to convince others, defend their standing points, and communicate their reasoning. These are often achieved through using argumentative devices~\cite{Walton2008, toulmin2003uses}. Take for example the discussion of Issue \#3064 of the Atom project\footnote{https://github.com/atom/atom/issues/3064}, an open-source text editor: A user opened the discussion by expressing frustration over the absence of a keyboard shortcut, arguing it as a common expectation for Linux users and a default on MacOS. After another user pointed out that there was indeed a hidden shortcut, the user who posted the issue argued that the shortcut should be visible and configurable. Then, different arguments were made by multiple users about whether to add the shortcut and what default key combinations it should be. Users laid out various types of reasoning to support their arguments, including (1) pointing out that the default key combination conflicts with another shortcut, (2) discussing the practices of other products and arguing for consistency, (3) emphasizing the importance of being unique, and (4) contrasting the needs of expert and general users. Eventually, the discussion led to a pull request that added the keyboard shortcut as a configurable menu item. As demonstrated in this example, exchanging opinions, posting statements, and supporting positions in the discussion are important ways the diverse OSS collaborators engage in discussions related to usability, share their points of view, and contribute to the ongoing improvements of OSS products. These argumentative discourses bring in viewpoints from different stakeholders and have the potential of serving as a source of ``collective intelligence''~\cite{Suran2020} that the OSS community can build on to better understand and satisfy the needs of their end users.

Previous work has investigated ways to leverage the argumentation structure in better supporting OSS community members to understand and consolidate usability discussions~\cite{wang2020argulens}. However, such techniques rely on high-quality arguments that people make, including the clarity of the claim, the credibility of the evidence, the overall persuasiveness, and more. Moreover, the quality of the arguments around usability highly affects how well the OSS community understands users' needs and addresses this important software aspect. In reality, however, many usability issue posters, often end users, do not have the ability to provide coherent and convincing reasonings~\cite{Wang2022IEEESoftware, Hellman2022Characterizing}. There is also very little support available to help OSS community members make strong arguments to raise awareness and attention around usability, which are already considered low priority~\cite{Wang2022IEEESoftware}. 

To provide meaningful support to OSS communities to effectively discuss usability-related topics, a comprehensive understanding is needed about the characteristics and quality of the arguments currently presented in these discussions. While previous studies have focused on argument components~\cite{dumani_systematic_2019, bilu_automatic_2015, Pang2008opinion, Liu2023} and argument quality dimensions~\cite{dumani_quality-aware_2020, wachsmuth_computational_2017} in other online forums, there is no synthesized knowledge about the characteristics of these aspects in the context of OSS usability and how they affect the discussion of usability-related topics. In this paper, we aim to address this gap. Particularly, we focus on achieving this through an analysis of OSS usability issue discussions on GitHub to answer the following research questions:

\begin{itemize}
    \item \textbf{RQ1}: How is the argumentative discourse expressed in the OSS usability issues?
    \item \textbf{RQ2}: What are the characteristics of the argument quality of the OSS usability issues?
    \item \textbf{RQ3}: To what extent does argumentation influence the discussions in usability issue threads?
\end{itemize}

To answer these research questions, we used an existing dataset about OSS usability discussions in five widely used OSS tools (Jupyter Lab, Google Colab, CoCalc, VSCode, and Atom)~\cite{sanei2023characterizing} and adopted a mixed-methods approach in our analysis. To characterize the argumentative discourse, we conducted a qualitative content analysis to detect the \textit{claims} and \textit{premises} in arguments and identified the types of claims and premises that OSS community members frequently use in their argument discourse. To characterize the argument quality, we adapted the dimensional framework introduced by \citet{wachsmuth_computational_2017}. This framework captures three main argument quality dimensions: \textit{logical} quality (or \textit{cogency}) indicating the argument's clarity and cohesiveness, \textit{rhetorical} quality (or \textit{effectiveness}) reflecting the argument's persuasive effect, and \textit{dialectical} quality (or \textit{reasonableness}) addressing the argument's strength in countering opposing viewpoints; each contains several sub-dimensions. We investigated how these dimensions manifested in the context of OSS usability issue discussions. Finally, to examine the impact of argumentation on the usability issue discussions, we employed statistical analysis to inquire the association between various argument dimensions and a set of usability issue discussion attributes related to the participants' collective behavior.

Our findings indicated that OSS contributors widely applied argument discourse in their usability discussions. Issue posts themselves were more likely to contain an argument than comments. The arguments presented in the issue posts also tended to be more complete and have a higher quality. Moreover, visual content, external files, and personal or hypothetical use cases used to support the arguments were associated with higher argument quality. Additionally, explicit premise in the argument of the issue posts encouraged more participants to contribute to the discussion, making a larger number of comments, while applying assertive claims led to quicker issue closure. Overall, our investigation painted a complex picture regarding the argument structure, quality, and their relationship with the usability issue discussion characteristics. Through this analysis, we derived insights to further enhance the quality of usability posts and to facilitate diverse participants to collaborate in OSS usability discussions. These insights can also inform other types of distributed, asynchronous collaborative work that involves diverse participants.

\section{Background and Related work}
This research is related to previous studies that focused on (1) usability issues in open-source software projects, (2) issue discussions in open-source software, and (3) understanding argumentation in open-source software projects.

\subsection{Usability Issues in Open Source Software Projects}
The practice of OSS has undergone remarkable growth and evolution over the past years. With the widespread of this type of software development model, concerns about usability of OSS have also increased~\cite{raza2012users,cheng2018open}. Usability encompasses various factors such as the effectiveness, efficiency, and satisfaction of users when interacting with a software system~\cite{nielsen1994usability}. It plays a pivotal role in the overall success and sustainability of OSS~\cite{nichols2006usability}. The significance of usability in OSS impacts the user experience, which, in turn, influences user adoption and the community's willingness to contribute to a project~\cite{Schwartz2009integrating}. Several studies have identified a range of usability challenges in OSS projects. These challenges often stem from the distributed, indirect collaboration and volunteer-driven nature of OSS~\cite{Wang2022IEEESoftware,nichols2006usability}. Moreover, developers who are inexperienced in understanding users and do not have enough knowledge of design constitute the majority of OSS contributors~\cite{Goldenvalue2005, raza2012users}. At the same time, users with diverse characteristics who may contribute to the improvement of usability, such as women~\cite{Trinkenreich2022}, users with disabilities~\cite{Aljedaani2024}, community-centric contributors~\cite{Trinkenreich2020}, and non-technical end users~\cite{Hellman2022Characterizing}, were often marginalized in OSS communities.

These challenges have prompted recent studies to examine how OSS developers address usability concerns. \citet{twidale2005exploring} investigated how OSS participants collaborate on usability issues with the help of issue tracking systems and identified several challenges related to the discussion dynamics and participant involvement. \citet{terry2008ingimp} presented the concept of open instrumentation, arguing that OSS projects should openly collect, make available, and learn from the rich usage data produced by end-users; they demonstrated this idea in \textit{ingimp}, a version of the OSS graphics editor that supports open instrumentation. A review of case studies about usability in eight Free/Libre/Open Source Software (FLOSS) projects also showed that an important problem regarding a usability initiative was the lack of user research~\cite{paul2009survey}. \citet{cetin2009collaboration} identified different types of collaboration methods among usability experts, developers, and end-users, focusing particularly on OSS projects, aiming to foster a better collaboration pattern. \citet{terry2010perceptions} also identified that the usability focus of OSS projects is positively associated with the size of user-base and a healthy relationship between the developers and the core users. \citet{cheng2018open} reported on an exploratory study that investigated how the OSS communities currently reported, discussed, negotiated, and eventually addressed usability and UX issues. Most recently, \citet{sanei2023characterizing} conducted a comprehensive analysis of OSS usability discussions and found that although usability issues were frequently discussed in OSS issue tracking systems, the scope of such discussion was limited to only a few usability aspects like efficiency and aesthetics. Their study has also resulted in a dataset of usability discussions that we leveraged in our work.

Extending this body of literature, in this research paper, we focus on understanding how OSS communities used arguments in usability issue discussions. While previous studies have highlighted the challenges and concerns within OSS usability discussions, our research concentrated on unraveling the argumentative discourse and quality embedded in these types of discussions, aiming to eventually providing insights to address those challenges and concerns.

\subsection{Issue Discussions in Open Source Software Projects}
Issue Tracking Systems (ITSs) are centralized, forum-like platforms for OSS communities to engage in multifaceted discussions around the development of their projects. Modern ITSs such as GitHub Issues~\cite{githubGitHubIssues} introduced extended collaboration features, such as comments, tags, and emoji reactions, to foster discussions on different software-related dimensions~\cite{skitalinskaya_learning_2021, dowden1993logical, Bertram2010}. Research into ITSs have examined diverse activities supported by issue discussions within OSS projects, including software requirements analysis~\cite{Heck2017}, bugs triaging~\cite{Xia2017}, features detection~\cite{Merten2016}, design rationales retrieval~\cite{Viviani2018}, and traceability enhancement~\cite{Nicholson2020}. Rich information is embedded in these discussion threads, used for exploring the causes of the issue, discussing the appropriateness and feasibility of proposed solutions, managing the project and the community, and engaging in social conversations, to name a few~\cite{arya2019analysis}. Therefore, many recent studies focused on understanding how OSS community collaborators interact in these discussion threads. For example,~\citet{Rath2020} introduced three interaction patterns in discussions: monologue, feedback, and collaboration. \citet{sanei2021impacts} found that emotional factors such as sentiments and tones impacted how OSS community members responded to the discussion. \citet{Li2021} found that codes of conduct were frequently used by the OSS community members and project maintainers to regulate community behavior. \citet{Gilmer2023} also investigated collaborative summarization techniques to support users' information acquisition and collective sense-making in issue discussions. Many recent studies have also investigated the complex phenomenon of uncivil and toxic behavior that occurred in ITSs and related software engineering artifacts~\cite{ferreira2021shut, Miller2022Toxicity, Qiu2022, Egelman2020}. For example, \citet{ferreira2021shut} found that two-thirds of non-technical code review discussions were uncivil and frequently exhibited features such as frustration, name calling, and impatience. \citet{Miller2022Toxicity} also identified that toxic behaviors observed in OSS communities displayed unique characteristics and were frequently triggered by difficulties using the software and ideology differences among the discussion participants. 

Our research builds upon and contributes to the literature that highlights the evolution of ITSs as a community-centric discussion platform. While previous studies identified patterns of interaction, emotional discussions, and diverse conversational dynamics, this study focused on a novel exploration of the argumentative discourse and quality of OSS issue discussions. This exploration enriches our understanding of the collaboration patterns of OSS communities on ITSs, particularly on usability issues.

\subsection{Understanding Argumentation in Open Source Software Projects}
Beyond the scope of OSS, there is a substantial body of literature focused on the related area of argumentation detection and analysis~\cite{PalauArgumentMining2009,mochales2011argumentation}, especially in online discussions. For example, \citet{cabrio2012combining} explored textual entailment for constructing argumentation networks and determining argument acceptability, emphasizing Dung's argumentation theory~\cite{DungTheory1995} and applying it to online debates. \citet{boltuvzic2014back} also introduced a supervised method for argument recognition with the help of a manually annotated corpus called ComARG, which worked according to comment-argument similarity, semantic comparisons, and textual entailment.

There are only a few prior studies focused on understanding the use of argumentation in the context of OSS. For example, the research of Yu et al.~\cite{Yu2011openargue} explored the application of argumentation analysis techniques in supporting software requirements discussions and developed OpenArgue; this tool facilitates syntax checking, argument reasoning, visual analysis of argument structures, and argument formalization. Most closely related to our study, Wang et al.~\cite{wang2020argulens} introduced a conceptual framework, ArguLens, leveraging Toulmin's argument model~\cite{toulmin2003uses} in supporting, understanding, and unifying opinions in ITSs.

While the previous works mostly focused on argument identification and classification, this research investigates how argumentative structure and quality manifest in and impact OSS discussions, outlining how argumentation plays a role in the collaborative efforts within OSS communities. Also, this study provided valuable insights into how argumentation can be used to improve user engagement and collaboration in OSS when addressing usability issues. Such insights can be leveraged to enhance other distributed, asynchronous collaborative work that involves diverse participants, helping them to articulate their viewpoints more effectively and fostering more inclusive collaboration environments.

\section{Methods}
\subsection{Dataset}
In this research, our focus was primarily on widely recognized and actively maintained OSS applications that have a Graphical User Interface (GUI); as a result of their popularity, they attracted a diverse community of members who participated in discussing usability concerns. Considering these criteria and our interest in investigating the spectrum of argument quality in usability issues, we used the labeled usability dataset from previous research~\cite{sanei2023characterizing}. The dataset was gathered in July 2021 and contained 127,282 issue discussions from five popular data science notebook and code editor projects hosted on GitHub. In the dataset, there is a random sample of 1,734 issues across the five projects manually labeled on (1) whether each issue touched upon usability concerns and, if so, (2) the main usability aspect it focused on, captured by Nielsen's usability heuristics~\cite{nielsen2005ten}. A total of 304 usability issues were identified this way. The five projects included in this dataset are:
\begin{itemize}
    \item Jupyter Lab\footnote{https://github.com/jupyterlab/jupyterlab} (88 usability issues, 405 comments in the dataset), which is a web-based interactive development environment for creating and sharing data science code and documents. This project started in 2016 and attracted more than 14,000 contributors.
    \item Google Colab\footnote{https://github.com/googlecolab/colabtools} (33 usability issues, 103 comments in the dataset), which is a cloud-based platform that allows users to work with data science notebooks using Google’s computing resources. This project was created in 2017 and attracted more than 19,000 contributors.
    \item CoCalc\footnote{https://github.com/sagemathinc/cocalc} (64 usability issues, 129 comments in the dataset), which is a cloud-based collaborative platform for data science and other computational projects, particularly for educational purposes. This is a relatively small project that was created in 2015 and involved around 50 contributors. 
    \item VSCode\footnote{https://github.com/microsoft/vscode} (82 usability issues, 333 comments in the dataset), which is a lightweight code editor developed by Microsoft that supports multiple programming languages and offers extensive customization through extensions. This project, started in 2015, had around 2,000 contributors.
    \item Atom\footnote{https://github.com/atom/atom} (37 usability issues, 174 comments in the dataset), which is a text and code editor developed by GitHub that features customizable interfaces and a rich ecosystem of plugins. This project was created in 2012 on GitHub and had around 500 contributors.
\end{itemize}

In addition to the issue and comment texts, the dataset also included information related to each issue, such as timestamps (posting time, closing time, time to first and last comment), the issue ID and URL, the number of comments, the number of participants in the discussion, and the number of reactions to the posted issue. In this study, we used the usability issues identified in this dataset to analyze the manifested argument discourse and quality.

\subsection{Characterizing Argumentative Discourse (RQ1)}
In this section, we present the details of the qualitative and quantitative analysis used to address RQ1, which is focused on identifying argumentative discourse in OSS usability issue discussions.

\subsubsection{Distinguishing argumentative discourse in usability discussion}
To investigate the characteristics of argumentative usability issues, we first coded the usability discussions in the dataset to differentiate the argumentative issues and comments from the non-argumentative ones. To this end, we focused on identifying argument \textit{claims} and \textit{premises}. We defined several criteria to distinguish claims and premises based on related literature~\cite{skitalinskaya_learning_2021,   wachsmuth_argumentation_2017, dowden1993logical}. Specifically, we considered a statement as a \textit{claim} if it (1) explicitly indicates a position or stance of the author regarding the discussed usability matters or (2) encapsulates the central point of view regarding usability aspects presented by the author. We considered a statement as a \textit{premise} if it (1) provides reasoning, evidence, example, or rationale supporting the main claim or stance presented by the author or (2) presents reasoning or grounds implicitly supporting a certain stance, even if the claim itself is not explicitly mentioned. We then considered an issue post or a comment to be argumentative if it contained a claim, a premise, or both; each issue report or comment could contain multiple arguments. Thus, an argument we identified can be one of the following three types, as argument structure: (1) \textit{claim only}, which only stated the stance without any support; (2) \textit{premise only}, which had implicit stance or claim with explicit supporting and reasoning, or (3) \textit{claim and premise}.

Subsequently, we conducted a qualitative content analysis~\cite{Drisko2016ContentAnalysis} to identify the types of claims and premises. For the claim types, we focused on identifying the types of statements that the claim used. While the coding process was essentially inductive, we were influenced by the terminology used by \citet{dowden1993logical} (i.e., proposition, assertion, judgment, hypothesis, principle, thesis, and law), since we were exposed to this framework before analysis. The coding process, however, started from the data and aimed at capturing the characteristics of usability issue discussions. For the premise types, we focused on identifying common themes in the resources the premises leveraged to support the corresponding claim. These themes were specific to usability issue discussions and were identified in an inductive manner.

The specific coding process is as follows. First, we randomly sampled five usability issue discussion threads in each project (in total 25, including their 113 comments), and two authors (first and fourth) independently coded the issue posts and comments according to the schema and focus described above. Then, they discussed their coding and generated a codebook that defined criteria for identifying claims, premises, the types of claims, and the types of premises. This round of annotating led to the presentation of more detailed and clear definitions to detect claims and premises. The first author then used this codebook to analyze the remaining 279 usability issues (including a total of 1,023 comments); during this process, any uncertainties in the coding or vagueness in the codebook were discussed among the authors and used to update the codebook.

To evaluate the reliability of the coding and further improve the codebook, we randomly sampled 20\% of all the usability issues (or 61 issues) that were coded by another researcher using the codebook and assessed the inter-rater reliability using Cohen's Kappa~\cite{Viera2005Kappa}. The inter-rater reliability was considered ``Substantial'' in all categories ($\kappa=0.77$, $0.76$, and $0.78$ for argument structure, claim type, and premise type, respectively). The two coders discussed their disagreements, reached a full agreement on the sampled data, and modified the codebook to provide further clarifications and guidance. The first author then coded all the usability issue discussion threads again using the updated codebook.

\subsubsection{Identifying the dominant argument in each argumentative issue/comment}
Each issue post or comment can contain more than one argument (in 403 issue posts and comments out of 1,095 argumentative data points in the dataset), making subsequent analysis unnecessarily complex. To address this, we aimed to identify the dominant argument in those issue posts and comments using the following two steps. First, in the issue post, if the claim is the title, then that claim and the corresponding premise (if any) constitute the dominant argument, since that claim is the dominant topic which the posted issue is focused on reporting. Then, we read each remaining issue post or comment and manually identified the dominant argument according to the context; we observed that the dominant argument is often associated with the first or the last claim of the issue post or comment.

\subsubsection{Association of usability dimensions with argument discourse}
We hypothesized that different usability dimensions (captured in Nielsen's usability heuristics~\cite{nielsen2005ten} in the dataset) might be associated with different claim types and premise types. For this analysis, only the six dominant usability dimensions labeled in the dataset were considered (i.e., \#7: Flexibility and efficiency of use, \#8: Aesthetic and minimalist design, \#9: Help users recognize, diagnose, and recover from errors, \#1: Visibility of system status, \#4: Consistency and standards, and \#5: Error prevention). These six dimensions covered more than 95\% of the discussions in the dataset and the remaining four dimensions, which only constituted an extremely small portion of the dataset (15 out of 304 issues), were disregarded to avoid causing bias due to the rare events. Because all variables are nominal types, we applied chi-squared tests to analyze the differences among the groups of usability dimension; the effect size of the correlation was further evaluated using Cramer's V. When a strong correlation was found, we conducted post hoc pairwise comparisons with Holm-Bonferroni correction to identify the pairs contributing to the difference. All the $\alpha$ level in all the statistic analyses of this paper was set at 0.05.

\subsection{Characterizing Argument Quality (RQ2)}
To address RQ2, we conducted the following qualitative and quantitative analyses. These investigations are centred on not only identifying argument quality dimensionality but also characterizing these dimensions in usability issue discussion threads.

\subsubsection{Assessing argument quality in usability discussions}
The aim of the qualitative analysis was to assess and detect the quality dimensions of arguments in the usability issues. For this, we adopted the taxonomy of argumentation quality proposed by~\citet{wachsmuth_computational_2017}. This taxonomy is by far the most extensive framework characterizing argument qualities. It is created by comprehensively synthesizing a large number of diverse theories and approaches focused on assessing argument quality. Overall, this taxonomy distinguished three main quality dimensions (logical, rhetorical, and dialectical qualities), each with specific sub-dimensions (see Table~\ref{tab:argument_quality_dimensions} for a summary). Each sub-dimension in this framework can be rated in three levels (i.e., high, medium, low), based on an annotation guideline accompanied the taxonomy~\cite{wachsmuth_computational_2017}.

\begin{table*}[t]
  \centering
  \small
  \caption{Summary of the taxonomy of argumentation quality proposed by ~\citet{wachsmuth_computational_2017}}
  \label{tab:argument_quality_dimensions}
  \resizebox{\textwidth}{!}{
      \begin{tabular}{lll}
        \toprule
        Dimension & Sub-dimension & Definition\\
        \midrule
        \multirow{3}{*}{Cogency (logical)} & Local acceptability & The premises are rational and reasonable.\\
         & Local relevance & The premises are relevant to the claim.\\
         & Local sufficiency & The premises are sufficient to support the claim.\\
        \midrule
        
        \multirow{5}{*}{Effectiveness (rhetorical)} & Credibility & The argument conveys that the author can be trusted.\\
         & Emotional Appeal & The argument creates emotions in favour of acceptance of its conclusion.\\
         & Clarity & The language used is correct, succinct, and unambiguous.\\
         & Appropriateness & The communication style is in favor of acceptance of the argument.\\
         & Arrangement & The language structure is in favor of acceptance of the argument.\\
        \midrule
    
        \multirow{3}{*}{Reasonableness (dialectic)} & Global acceptability & The argument is likely to be accepted by readers.\\
         & Global relevance & The argument provides valuable solutions or insights.\\
         & Global sufficiency & Counter-arguments are sufficiently rebutted.\\
        \bottomrule
      \end{tabular}
  }
\end{table*}

To explore the applicability of this taxonomy to usability issue arguments, two authors (first and fourth) first coded the argument quality on 25 usability issues and their 113 comments (the same random sample used for developing the codebook in RQ1). In this step, the two authors focused on adapting the argument quality definitions of~\citet{wachsmuth_computational_2017} in assessing the quality of usability-related arguments. They noted their coding rationale on the three levels for each sub-dimension, capturing the specific characteristics of usability issue discussions, and created an initial codebook. Then, the first author used this codebook to analyze the argument quality of (1) all 285 issue posts that included an argument and (2) 351 comments on the five longest argumentative issues in our dataset for each project. The decision of only focusing on the longest issues for analyzing argument quality in comments was made to manage the extensive manual effort required for this analysis while capturing the nuances and complexity of issue discussions, following the same rationale in the literature~\cite{wang2020argulens, arya2019analysis}; the median length of the longest issues were $M=17$ ($total=84$ comments) for Jupyter Lab, $M=6$ ($total=54$) for Google Colab, $M=6$ ($total=29$) for CoCalc, $M=18$ ($total=103$) for VSCode, and $M=13$ ($total=81$) for Atom.
The codebook was refined and enriched during this process.

After this step, an additional researcher, who was involved in the first stage of qualitative analysis of argument discourse, applied the argument quality codebook on the sampled dataset of 61 issues (20\% of all the usability issues) to assess the inter-rater reliability using Cohen's Kappa. The average Kappa value on each sub-dimension of argument quality was $\kappa= 0.77$ ($SD=0.12$), indicating a substantial agreement. The two coders also discussed their disagreements, leading to slight refinements in the quality dimensions codebook.

\subsubsection{Association of usability dimensions and argument discourse with argument quality.} 
To investigate the association of usability dimensions, claim types, and premise types of the posted issues as independent variables (nominal data type) with the argument quality dimensions as dependent variables (ordinal data type), we first performed Kruskal-Wallis tests and calculated the effect size with $\eta^2$. If a strong correlation was found, we performed post hoc pairwise tests with Holm-Bonferroni correction to distinguish the significant pairs. For the usability dimensions, we again considered the six most frequently appeared heuristics in the dataset. The $\alpha$ level was set at 0.05.

\subsection{Impacts of Argument on Usability Discussion Threads (RQ3)}
To address RQ3, we conducted a series of quantitative statistical analyses on the annotated dataset of each project. Particularly, we concentrated on examining a set of attributes related to the behavior of participants in posing argumentative posts (issues or comments) of usability. These attributes included: 

\begin{itemize}
    \item \textbf{Discussion length}: This attribute represents the number of comments posted for a usability issue, indicating the complexity of the issue and the richness of the discussion.
    \item \textbf{Number of participants}: This attribute represents the number of unique participants who contributed to the discussion, indicating participant involvement.
    \item \textbf{Time to first comment}: This attribute represents the time span from the issue opening to the first comment, indicating how fast other collaborators join the discussion and respond to the issue.
    \item \textbf{Time to close the issue}: This attribute represents the time span from the issue opening to the end of the discussion, indicating how long it took to complete an issue. 
    \item \textbf{Number of reactions}: GitHub allows its community members to make emoji reactions such as thumbing up, thumbing down, smiley face, etc. This attribute captures the number of reactions the community made to the usability issue, indicating how much this issue attracted the attention of community members.
\end{itemize}

We applied statistical analyses to investigate the impact of a set of independent variables: (1) argument structure (nominal type), (2) claim types (nominal type), (3) premise types  (nominal type), and (3) argument quality dimensions (ordinal type) on the above attributes as dependent variables (all are interval data type). We initially performed the Shapiro-Wilk tests and confirmed that all dependent variables followed a normal distribution. As a result, an ANOVA test was executed to signify potential disparities within the groups of each independent variable. Then, pairwise post hoc analyses were performed with Holm-Bonferroni correction to identify the independent group pairs contributing to the difference. Again, the $\alpha$ level was set at 0.05.

\section{RQ1 Results: Argumentative Discourse in Usability Issues}

\subsection{Frequency of Argumentative Usability Issue Posts and Comments}

\begin{figure}[t]
  \centering
  \begin{minipage}
    [b]{0.48\textwidth}
    \includegraphics[width=\textwidth]{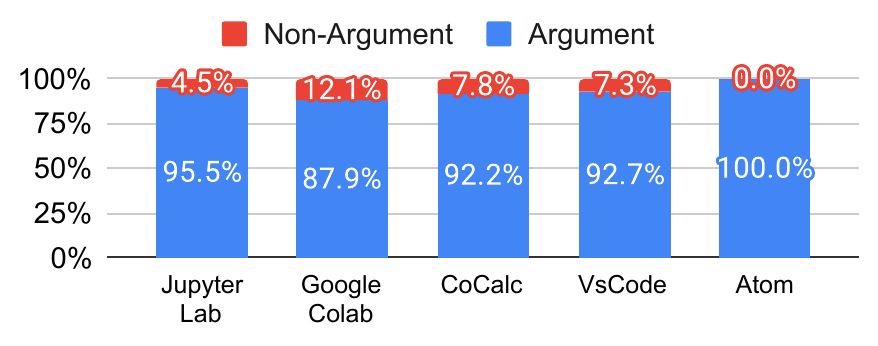}
   \subcaption{Issue posts}
    \label{fig:Argumentative-NonArgumentative-Issue}
  \end{minipage}
  \hfill
  \begin{minipage}[b]{0.48\textwidth}
    \includegraphics[width=\textwidth]{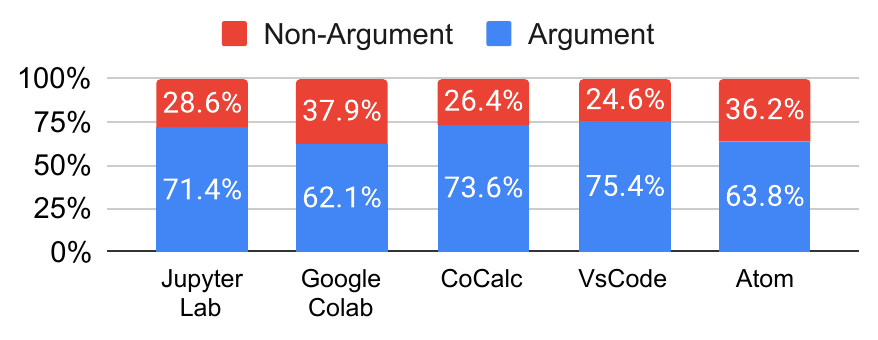}
    \subcaption{Comments} 
    \label{fig:Argumentative-NonArgumentative-Comment}
  \end{minipage}
   \caption{Frequency of argumentative and non-argumentative issue posts and comments in each project}
   \Description{Two bar charts are shown. The first one shows the percentages of argument and non-argument issue posts; the second shows percentages of argument and non-argument comments. For issue posts, the percentages of argumentative issues were 95.5\% in Jupyter Lab, 87.9\% in Google Colab, 92.2\% in CoCalc, 92.7\% in VSCode, and 100\% in Atom. For comments,  the percentages of argumentative issues were 71.4\% in Jupter Lab, 62.1\% in Google Colab, 73.6\% in CoCalc, 75.4\% in VSCode, and 63.8\% in Atom.}
   \label{fig:Argumentative-NonArgumentative}
\end{figure}

\begin{figure}[t]
  \centering
  \begin{minipage}
    [b]{0.48\textwidth}
    \includegraphics[width=\textwidth]{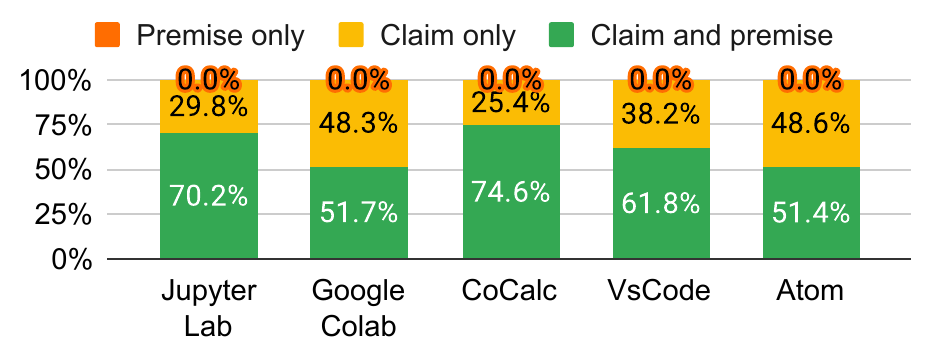}
   \subcaption{Issue posts}
    \label{fig:Argumentative-Type-Issue}
  \end{minipage}
  \hfill
  \begin{minipage}[b]{0.48\textwidth}
    \includegraphics[width=\textwidth]{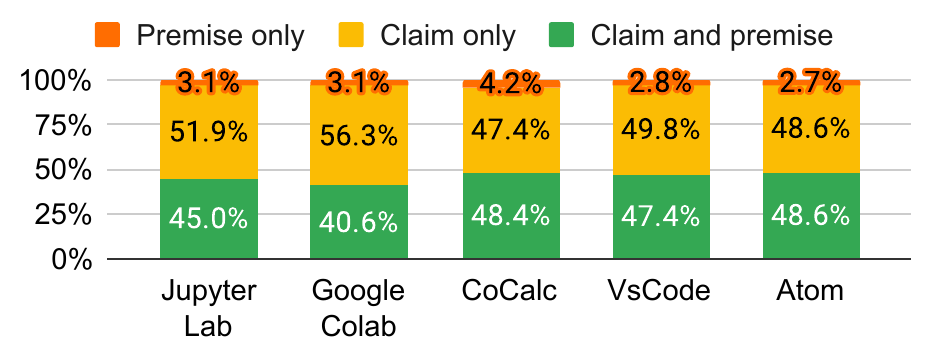}
    \subcaption{Comments} 
    \label{fig:Argumentative-Type-Comment}
  \end{minipage}
   \caption{Frequency of the three types of argument in issue posts and comments in each project}
    \Description{Two bar charts report the frequency of the three argument structure types in issue posts and comments. The first bar chart displays the percentage of argument structure in posted issues; the second one reports the percentages of argument structure in comments. For issue posts, the percentages of claim and premise arguments were 70.2\% in Jupyter Lab, 51.7\% in Google Colab, 74.6\% in CoCalc, 61.8\% in VSCode, and 51.4\% in Atom; the percentages for claim only arguments were 29.8\% in Jupter Lab, 48.3\% in Google Colab, 25.4\% in CoCalc, 38.2\% in VSCode, and 48.6\% in Atom; the percentages of premise only arguments in all the projects is 0\%. For comments, the percentages of claim and premise arguments were 45.0\% in Jupyter lab, 40.6\% in Google Colab, 48.4\% in CoCalc, 47.4\% in VSCode, and 48.6\% in Atom; the percentages of claim only arguments were 51.9\% in Jupter Lab, 56.3\% in Google Colab, 47.4\% in CoCalc, 49.8\% in VSCode, and 48.6\% in Atom; the percentages of premise only arguments were 3.1\% in Jupter Lab, 3.1\% in Google Colab, 4.2\% in CoCalc, 2.8\% in VSCode, and 2.7\% in Atom.}
    \label{fig:ArgumentType}
\end{figure}

We found that the majority of the usability issues were posted with an argument ($mean=93.7\%$ across the five projects, $SD=4.5\%$); see Figure~\ref{fig:Argumentative-NonArgumentative-Issue}. This percentage is lower in comments on usability issues, but still, about 2/3 of the comments were created with an argument ($mean=69.3\%$ across the five projects, $SD=6.0\%$); see Figure~\ref{fig:Argumentative-NonArgumentative-Comment}. Additionally, in issue posts, the majority of the arguments were created with both claim and premise ($mean=61.9\%$ across the five projects, $SD=10.5\%$); see Figure~\ref{fig:Argumentative-Type-Issue}. This percentage is lower in argumentative comments on usability issues ($mean=46.0\%$, $SD=3.3\%$); see Figure~\ref{fig:Argumentative-Type-Comment}. In issue posts, all arguments were made with a claim, while in comments, there are very few arguments that contained \textit{premise only}.

\subsection{Claim Types in Usability Arguments}

Through the inductive coding process, we identified the following four types of claims that users made in usability issues posts and comments:

\begin{itemize}[leftmargin=*]
    \item \textbf{Assertion:} In this type of claim, the participants declared their stance or position in a confident, certain, and self-assured style, and they meant exactly what they wished to convey. For example, in Atom issue \#3064, a poster made an assertive claim:  ``\textit{It's not displayed.}'' As another more extensive example, someone wrote assertively in Atom issue \#3512: ``\textit{This issue is not about performance, it's just about characters not included in highlighting.}
    
    \item \textbf{Hypothesis/opinion:} This type of claim stated the authors' position as something they thought, believed, preferred, had a view or perspective on, or had a theory or assumption about. For instance, a poster indicated personal belief in Atom issue \#5497: ``\textit{I believe that this should be documented below the checkbox in preferences.}'' Also, in Jupyter lab issue \#3901, an author wrote about their preferred solution to the issue: ``\textit{I'd prefer to just make it a user setting.}''
    
    \item \textbf{Probability/doubtfulness:} We coded this type of claim when the author was uncertain or skeptical about things they were discussing. For example, someone made a claim in Atom issue \#902: ``\textit{This probably won't be an issue with the new trapezoidal tabs.}'' Also, in VSCode issue \#5497, someone claimed that there ``\textit{seems to be some underlying code that causes [this issue]}''; using the word ``seems'' indicated the author's uncertainty and speculation.
    
    \item \textbf{Suggestion/recommendation:} We coded this type of claim when the author indicated their statement as advice, a proposal, or a tip to make an improvement. For example, someone claimed in CoCalc issue \#1238: ``\textit{we should implement drag and drop of image files onto chat messages}'', suggesting a recommended action for improving the chat feature. Also, in Jupyter issue \#1084,  the statement, ``\textit{it would be even better if this changed the theme}'', presents a specific suggestion.
\end{itemize}

\begin{figure}[t]
  \centering
  \begin{minipage}
    [b]{0.48\textwidth}
    \includegraphics[width=\textwidth]{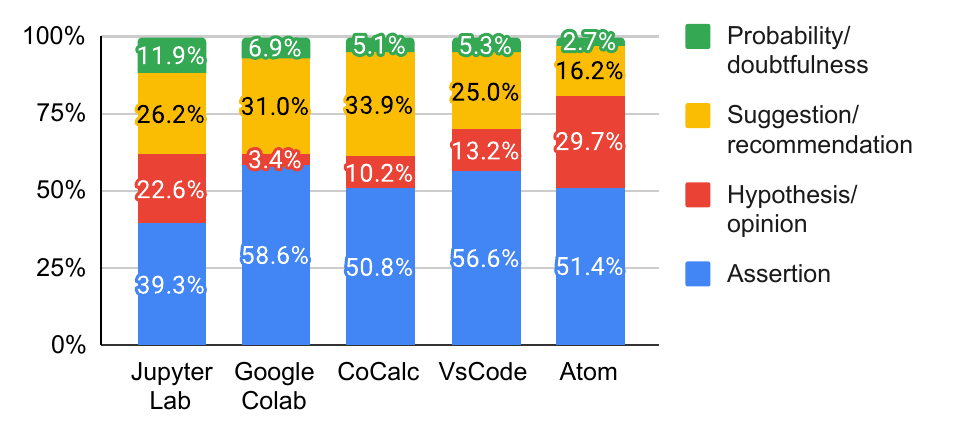}
   \subcaption{Issue posts}
    \label{fig:Claim-Type-Issue}
  \end{minipage}
  \hfill
  \begin{minipage}[b]{0.48\textwidth}
    \includegraphics[width=\textwidth]{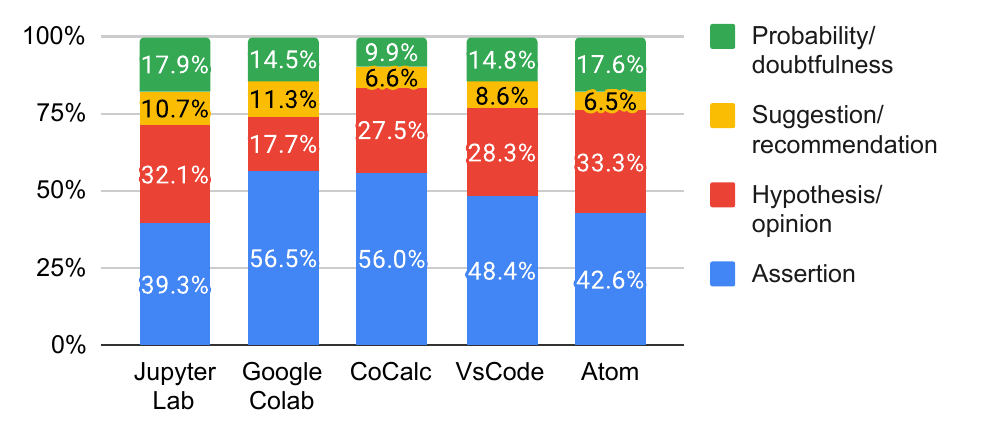}
    \subcaption{Comments} 
    \label{fig:Claim-Type-Comment}
  \end{minipage}
   \caption{Percentage of each claim type in issue posts and comments in every project}
    \Description{Two bar charts report the frequency of the four argument claim types in issue posts and comments. The first bar chart displays the percentage of argument structure in posted issues; the second one reports the percentages of argument structure in comments. For issue posts, the percentages of assertion claims were 39.3\% in Jupyter Lab, 58.6\% in Google Colab, 50.8\% in CoCalc, 56.6\% in VSCode, and 51.4\% in Atom; the percentages of hypothesis/opinion claims were 22.6\% in Jupter Lab, 3.4\% in Google Colab, 10.2\% in CoCalc, 13.2\% in VSCode, and 29.7\% in Atom; the percentages of suggestion/recommendation claims were 26.2\% in Jupter Lab, 31.0\% in Google Colab, 33.9\% in CoCalc, 25.0\% in VSCode, and 16.2\% in Atom; the rest is the percentages of probability/doubtfulness claims, ranging from 11.9\% to 2.7\% across the projects. For comments, the percentages of assertion claims were 39.3\% in Jupyter Lab, 56.5\% in Google Colab, 56.0\% in CoCalc, 48.4\% in VSCode, and 42.6\% in Atom; the percentages of hypothesis/opinion claims were 32.1\% in Jupter Lab, 17.7\% in Google Colab, 27.5\% in CoCalc, 28.3\% in VSCode, and 33.3\% in Atom; the percentages of probability/doubtfulness claims were 17.9\% in Jupter Lab, 14.5\% in Google Colab, 9.9\% in CoCalc, 14.8\% in VSCode, and 17.6\% in Atom; the rest is the percentages of suggestion/recommendation claims, ranging from 11.3\% to 6.5\% across the projects.}
   \label{fig:Claim-Type}
\end{figure}

We found that in both argumentative issue posts and argumentative comments, the most frequently used claim type was \textit{Assertion} ($mean=51.3\%$ $SD=7.5\%$ for issue posts and $mean=48.6\%$ $SD=7.7\%$ for comments); see Figure~\ref{fig:Claim-Type}. However, the second frequently used claim type differed between argumentative issue posts and argumentative comments. For most projects (except Atom), the second frequent claim type in issue posts was \textit{Suggestion/recommendation} ($mean=26.5\%$ $SD=6.8\%$), but that in argumentative comments was \textit{Hypothetical/opinion} ($mean=27.8\%$ $SD=6.1\%$). Furthermore, the expression of uncertainty and skepticism (with \textit{Probability/doubtfulness} claims) was the least common across all projects in issue posts; in contrast, the least common claim type in argumentative comments was \textit{Suggestion/recommendation}. We speculate that these differences reflected the different natures and goals of issue posts (i.e., describing a usability problem and proposing a solution to improve usability) and comments (i.e., expressing personal opinions or reactions related to the usability problem).

\subsection{Premise Types in Usability Arguments}

The qualitative analysis process allowed us to identify the following seven premise types in our dataset:
\begin{itemize}[leftmargin=*]
    \item \textbf{Visual content or supporting file:} The type of premise used external link, file, image, video, or code snippet to substantiate and support the claim. For example, in Google Colab issue \#11, a poster supported their point by writing: ``\textit{More info on markdown in colab is in [Link to File].}'' In VSCode issue \#116061, someone also used an image to provide context for their claim: ``\textit{For context, here is the type of error pop-up I'm talking about: [image].}''
    
    \item \textbf{Specific usage experience:} Premises of this type describe personal use experiences or hypothetical use cases embedded with a usage experience. Often, these experiences expressed a desire to be convenient, efficient, and/or less restrained when using the application. Sometimes, they provided evidence to support the claim that the software does not function as expected. For example, a poster in VSCode issue \#82247 described a use scenario to support their claim: ``\textit{Sometimes it is shown (a blue border) when I click the button, but then clicking outside the search widget, the icon is then greyed out... It's all a bit confusing.}'' Another example of this can be seen in VSCode issue \#45445, in which a poster supported their claim with a usage experience: ``\textit{because Find in selection is too unpredictable and causes time waste.}''
    
    \item \textbf{Evidence that the issue is or can be resolved:} This type of premise indicated or demonstrated that the issue is resolved or presented promising suggestions leading to resolving the issue. For instance, a poster in Atom issue \#902 indicated that no further discussion is needed since ``\textit{tabs were redesigned and shipped.}'' Also, in Atom issue \#11448, someone claimed that the issue can be closed ``\textit{Since this has been addressed in the One theme.}''
    
    \item \textbf{Comparing with competitors/consistency:} In this premise type, the author took into account similarities with other tools and/or aimed to maintain consistency with similar products while considering being aligned with the current software. For example, in Jupyter Lab issue \#1084, a poster justified their argument by writing: ``\textit{[Because] Sublime has a color scheme for the text area and a theme for UI itself.}''
    
    \item \textbf{Referring to another issue:} Premise of this type used another issue to support the writers' stance, sometimes indicating that the current issue is linked to another issue, resolved by another issue, or warrants opening another issue. For example, in CoCalc issue \#274, someone used the following premise to support their decision to close the issue: ``\textit{Close in favour of this issue: [link].}''
    
    \item \textbf{Clarification and providing additional information:} This premise type focused on providing factual details or explanations to enhance understanding or providing more context about a particular claim in order to support it. For example, in Atom \#3064, an author justified their argument of using a particular character for a keyboard shortcut by clarifying: ``\textit{because opening a file or a folder in dev mode use the same character.}''
\end{itemize}

\begin{figure}[t]
  \centering
  \begin{minipage}
    [b]{0.48\textwidth}
    \includegraphics[width=\textwidth]{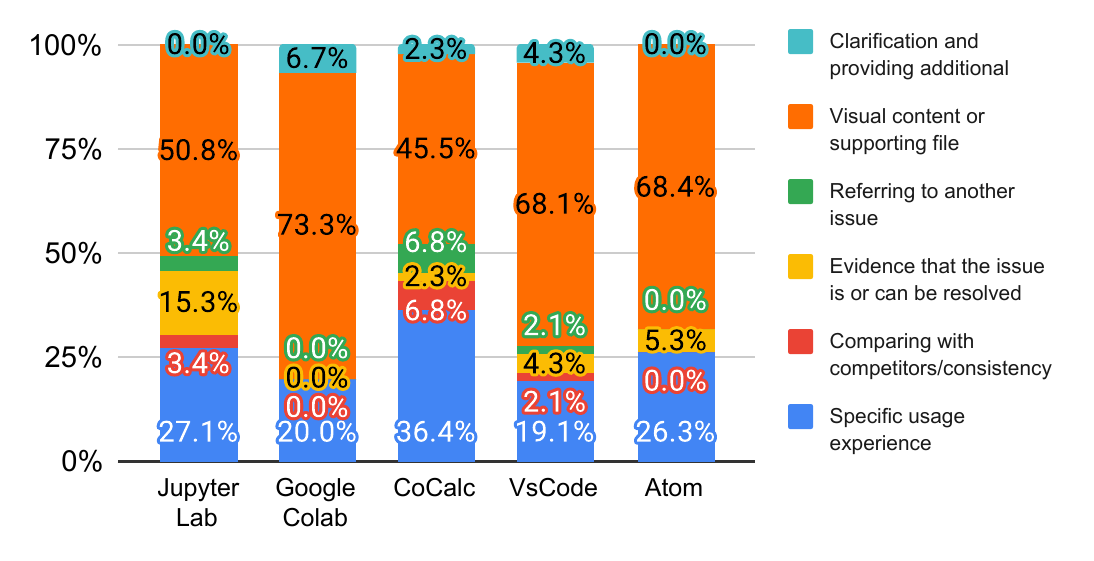}
   \subcaption{Issue posts}
    \label{fig:Premise-Type-Issue}
  \end{minipage}
  \hfill
  \begin{minipage}[b]{0.48\textwidth}
    \includegraphics[width=\textwidth]{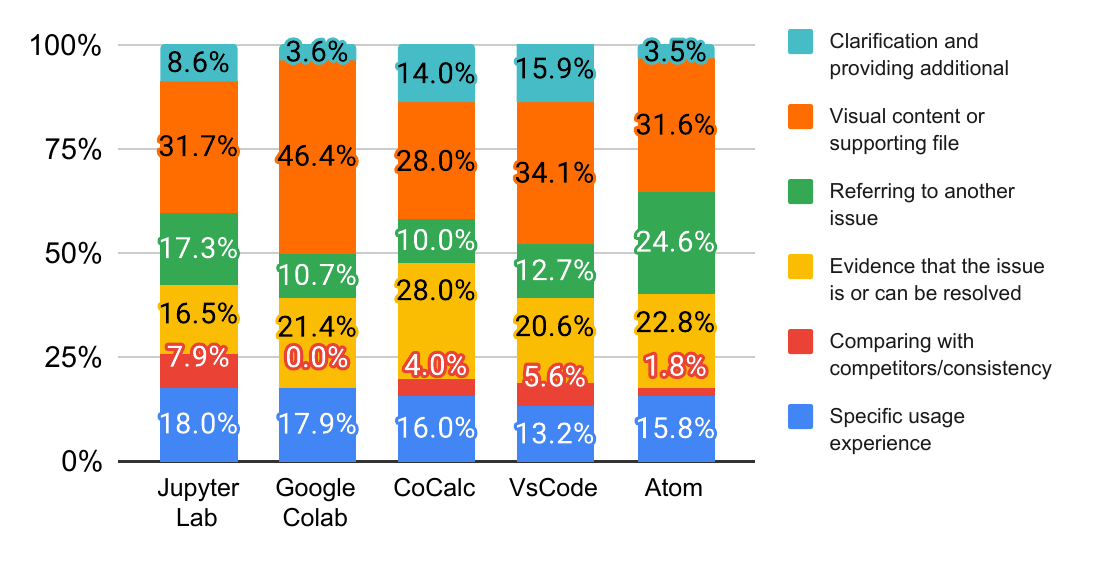}
    \subcaption{Comments} 
    \label{fig:Premise-Type-Comment}
  \end{minipage}
   \caption{Percentage of each premise type in issue posts and comments in every project}
   \Description{Two bar charts report the frequency of the six argument premise types in issue posts and comments. The first bar chart displays the percentage of argument structure in posted issues; the second one reports the percentages of argument structure in comments. For issue posts, the percentages of visual content and supporting file premises were 50.8\% in Jupyter Lab, 73.3\% in Google Colab, 45.5\% in CoCalc, 68.1\% in VSCode, and 68.4\% in Atom; the percentages of specific usage scenario premises were 27.1\% in Jupter Lab, 20.0\% in Google Colab, 36.4\% in CoCalc, 19.1\% in VSCode, and 26.3\% in Atom; the percentages of other premise types are small and all less than 10\%. For comments, the percentages of visual content and supporting file premises ranged from 46.4\% to 28.0\% across the projects; the percentages of evidence that the issue is or can be resolved premises ranged from 28.0\% to 16.5\%; the percentages of referring to another issue premises ranged from 24.6\% to 10.0\%; the percentages of specific usage scenario premises ranged from 18.0\% to 13.2\%; the percentages of clarification and providing additional information premises ranged from 15.9\% to 3.5\%.}
   \label{fig:Premise-Type}
\end{figure}

These premise types have different frequencies in each project for issues and comments separately (see Figure~\ref{fig:Premise-Type}); these frequencies also varied slightly across the five projects. In the issue posts, the most frequent premise types were \textit{Visual content or supporting file} ($mean=61.2\%$, $SD=12.2\%$) and \textit{Specific usage experience} ($mean=25.8\%$, $SD=6.9\%$). In the comments, there were several frequently appeared premise types, including \textit{Visual content or supporting file} ($mean=34.4\%$, $SD=7.1\%$), \textit{Evidence that the issue is or can be resolved} ($mean=21.9\%$, $SD=4.2\%$), \textit{Specific usage experience} ($mean=16.2\%$, $SD=1.9\%$), and \textit{Referring to another issue} ($mean=15.1\%$, $SD=6.0\%$). The other types of premise are relatively rare. Chi-square tests did not find a significant correlation among claim types and premise types.

\subsection{Association of Usability Dimensions With Argument Discourse}
We found that the type of usability issues (captured by Nielsen's heuristics) strongly correlated with the claim types of posted issues in two projects, CoCalc ($Cramer's V=0.361$, $p< 0.01$) and VSCode ($Cramer's V=0.241$, $p< 0.05$). The post hoc analysis revealed that, in both projects, usability arguments of issues related to the \textit{\#8: Aesthetic and minimalist design} heuristic were more frequently claimed with an \textit{Assertion}, while those related to \textit{\#7: Flexibility and efficiency of use} were more frequently claimed with a \textit{Suggestion or recommendation}. On the other hand, the only project in which we found a correlation between usability issue types and premise types was Atom ($Cramer's V=0.384$, $p< 0.05$). The post hoc analysis revealed that usability arguments related to \textit{\#7: Flexibility and efficiency of use} of issues were more likely not to include a premise.

\vspace{4pt}
\begin{tcolorbox}[colframe=black,colback=gray!10,boxrule=0.5pt,arc=.3em,boxsep=-1mm]
\textbf{RQ1 Main Findings}: The majority of the usability issues and their comments were posted with an argument. Arguments in issue posts were more likely to be supported by a premise than in comments. About half of the claims in the usability issues were \textit{Assertions}, and the most frequently used types of premise were \textit{Visual content or supporting file} and \textit{Specific usage experience}. Correlations between usability issue types and the claim and premise types were found in a few projects.
\end{tcolorbox}

\section{RQ2 Results: Argument Quality in Usability Issues}

\subsection{Argument Quality in Issue Posts and Issue Comments}
\label{sec:RQ2Results_1}

\begin{table*}[t]
  \centering
  \caption{Percentage of argument quality ratings in usability issue posts}
  \label{tab:argument_quality_issues}
  \resizebox{\textwidth}{!}{
  \begin{tabular}{@{}lccccccccccccccc@{}}
    \toprule
    & \multicolumn{3}{c}{Jupyter Lab} & \multicolumn{3}{c}{Google Colab} & \multicolumn{3}{c}{CoCalc} & \multicolumn{3}{c}{VSCode} & \multicolumn{3}{c}{Atom} \\
    \cmidrule(lr){2-4} \cmidrule(lr){5-7} \cmidrule(lr){8-10} \cmidrule(lr){11-13} \cmidrule(lr){14-16}
    & High & Med. & Low & High & Med. & Low & High & Med. & Low & High & Med. & Low & High & Med. & Low \\
    \midrule
Local Acceptability&	53.6 & 32.1 & 14.3 & 48.3 & 20.7 & 31.0 & 64.9 & 17.5 & 17.5 & 66.2 & 11.7 & 22.1 & 54.1 & 8.1 & 37.8 \\ 
Local Relevance&	52.4 & 33.3 & 14.3 & 51.7 & 13.8 & 34.5 & 63.2 & 19.3 & 17.5 & 68.8 & 7.8 & 23.4 & 51.4 & 10.8 & 37.8 \\ 
Local Sufficiency	&51.2 & 17.9 & 31.0 & 51.7 & 10.3 & 37.9 & 61.4 & 19.3 & 19.3 & 67.5 & 7.8 & 24.7 & 45.9 & 10.8 & 43.2 \\
\rowcolor{lavender}
    \textbf{Cogency} &	50.0 & 35.7 & 14.3 & 51.7 & 13.8 & 34.5 & 64.9 & 17.5 & 17.5 & 67.5 & 9.1 & 23.4 & 45.9 & 16.2 & 37.8 \\\midrule
Credibility&	23.8 & 76.2 & 0.0 & 10.3 & 89.7 & 0.0 & 22.8 & 77.2 & 0.0 & 18.2 & 81.8 & 0.0 & 13.5 & 86.5 & 0.0 \\ 
Emotional Appeal&	40.5 & 59.5 & 0.0 & 31.0 & 69.0 & 0.0 & 28.1 & 71.9 & 0.0 & 44.2 & 55.8 & 0.0 & 18.9 & 81.1 & 0.0 \\ 
Clarity	&26.2 & 72.6 & 1.2 & 37.9 & 62.1 & 0.0 & 36.8 & 63.2 & 0.0 & 40.3 & 59.7 & 0.0 & 16.2 & 81.1 & 2.7 \\ 
Appropriateness&	22.6 & 77.4 & 0.0 & 31.0 & 65.5 & 3.4 & 14.0 & 84.2 & 1.8 & 16.9 & 76.6 & 6.5 & 16.2 & 78.4 & 5.4 \\ 
Arrangement&	34.5 & 58.3 & 7.1 & 31.0 & 37.9 & 31.0 & 35.1 & 45.6 & 19.3 & 42.9 & 41.6 & 15.6 & 27.0 & 40.5 & 32.4 \\ 
\rowcolor{lavender}
    \textbf{Effectiveness} &	21.4 & 78.6 & 0.0 & 17.2 & 82.8 & 0.0 & 17.5 & 82.5 & 0.0 & 24.7 & 75.3 & 0.0 & 8.1 & 91.9 & 0.0 \\ \midrule
Global Acceptability	&56.0 & 41.7 & 2.4 & 55.2 & 20.7 & 24.1 & 59.6 & 28.1 & 12.3 & 67.5 & 22.1 & 10.4 & 54.1 & 24.3 & 21.6 \\ 
Global Relevance	&27.4 & 64.3 & 8.3 & 20.7 & 37.9 & 41.4 & 29.8 & 47.4 & 22.8 & 28.6 & 48.1 & 23.4 & 21.6 & 43.2 & 35.1  \\ 
Global Sufficiency	&38.1 & 42.9 & 19.0 & 44.8 & 20.7 & 34.5 & 35.1 & 42.1 & 22.8 & 50.6 & 32.5 & 16.9 & 43.2 & 24.3 & 32.4 \\ 
\rowcolor{lavender}
    \textbf{Reasonableness} &38.1 & 58.3 & 3.6 & 37.9 & 34.5 & 27.6 & 43.9 & 40.4 & 15.8 & 53.2 & 33.8 & 13.0 & 37.8 & 37.8 & 24.3 \\
    \bottomrule
  \end{tabular}}
\end{table*}

\begin{figure*}[t]
    \begin{subfigure}{.19\textwidth}
      \includegraphics[width=.99\linewidth]{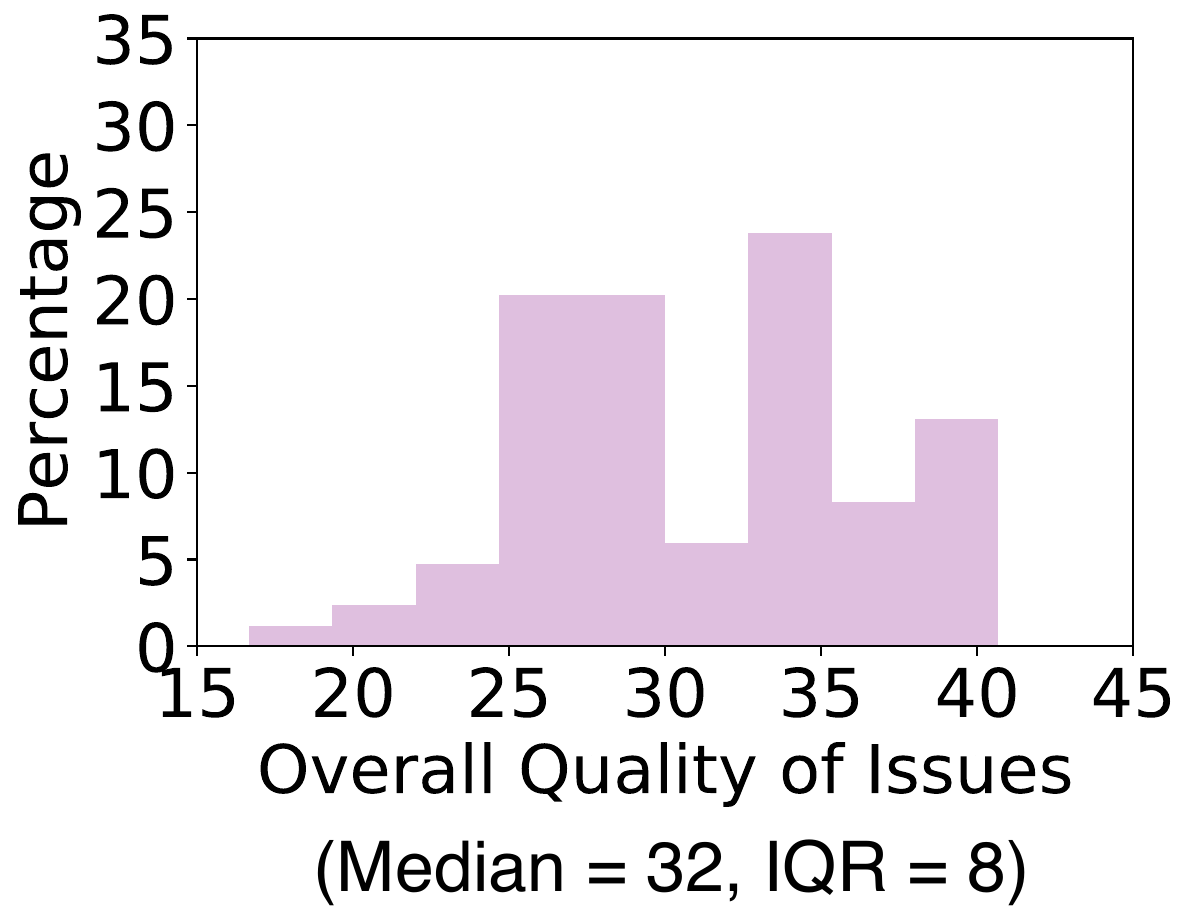}
      \caption{Jupyter Lab}
      \label{fig:overall_argument_quality_issues_jupyterlab}
    \end{subfigure}
    \begin{subfigure}{.19\textwidth}
      \centering
      \includegraphics[width=.99\linewidth]{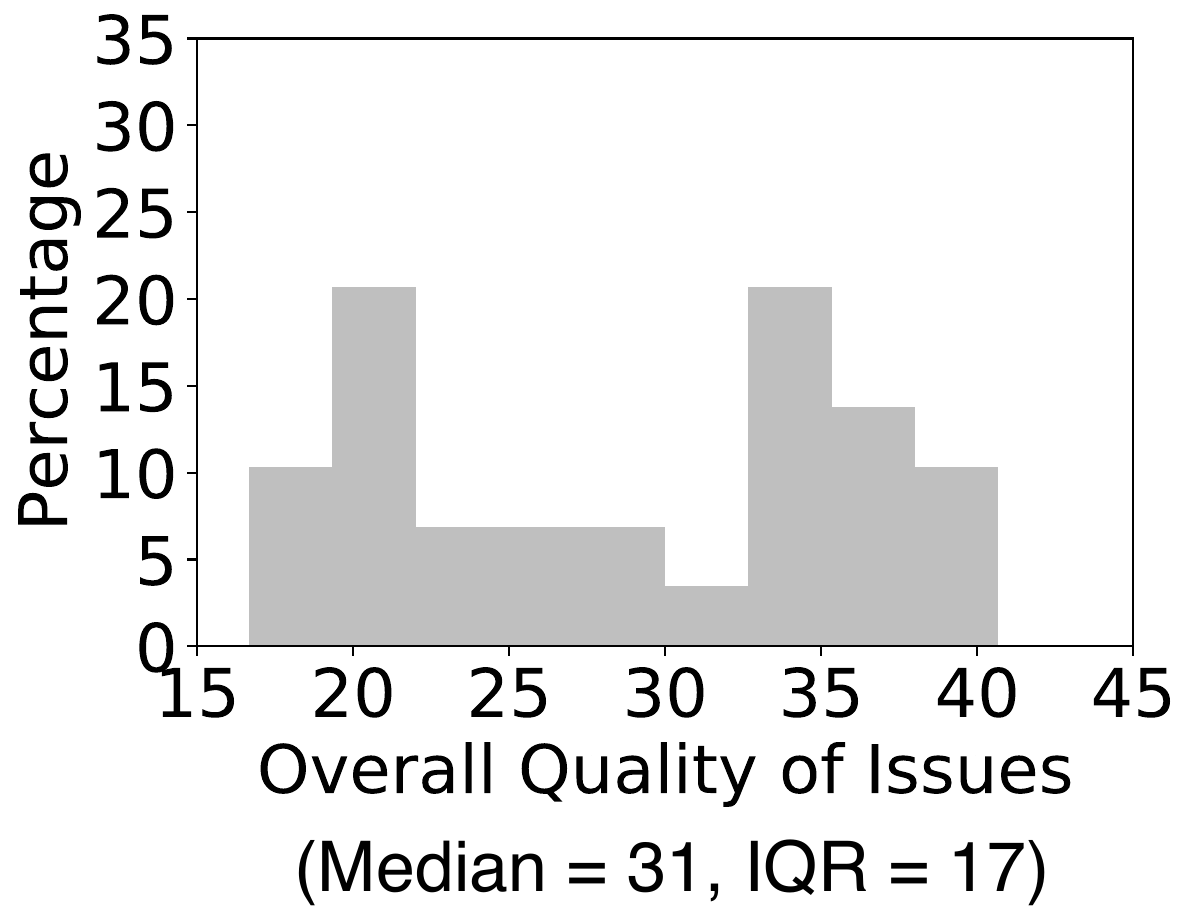}
      \caption{Google Colab}
      \label{fig:overall_argument_quality_issues_colab}
    \end{subfigure}
    \begin{subfigure}{.19\textwidth}
      \centering
      \includegraphics[width=.99\linewidth]{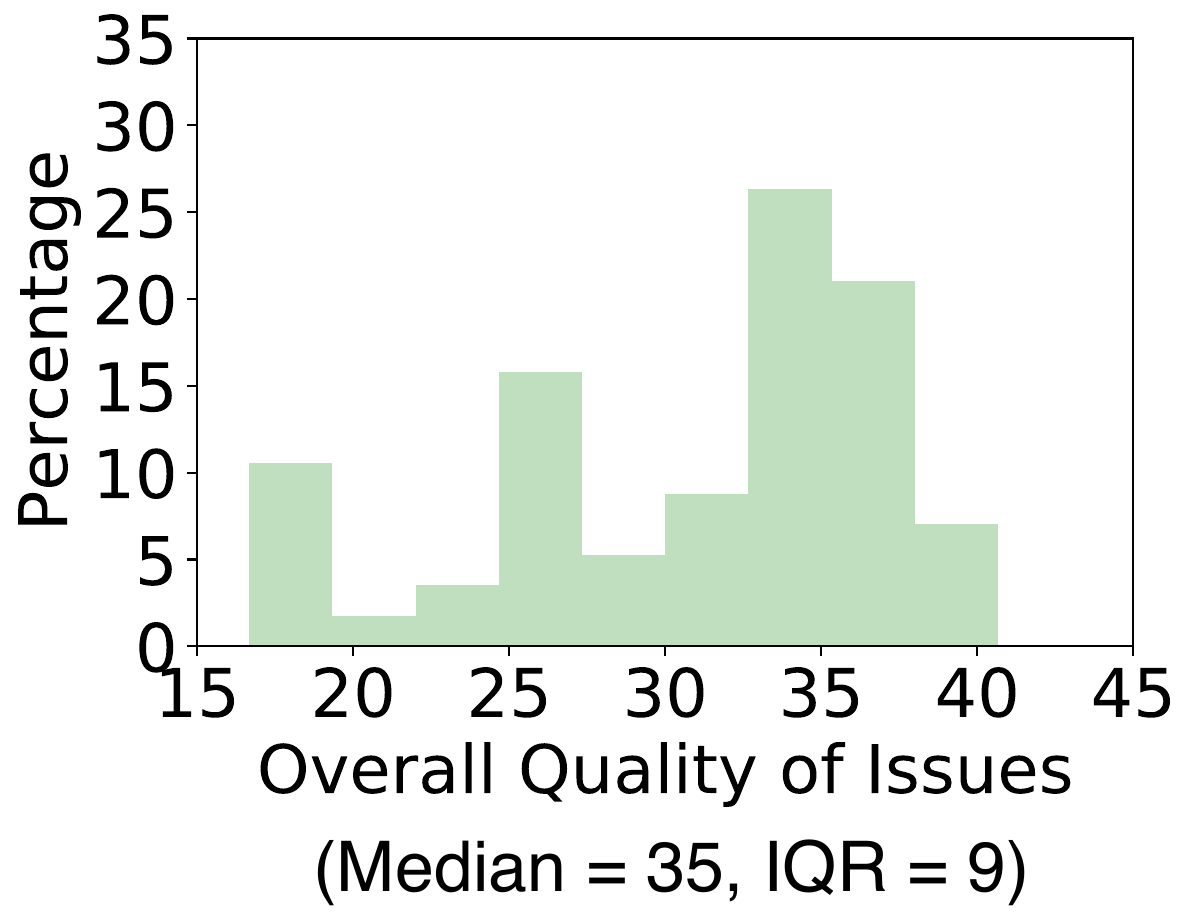}
      \caption{CoCalc}
      \label{fig:overall_argument_quality_issues_cocalc}
    \end{subfigure}
    \begin{subfigure}{.19\textwidth}
      \centering
      \includegraphics[width=.99\linewidth]{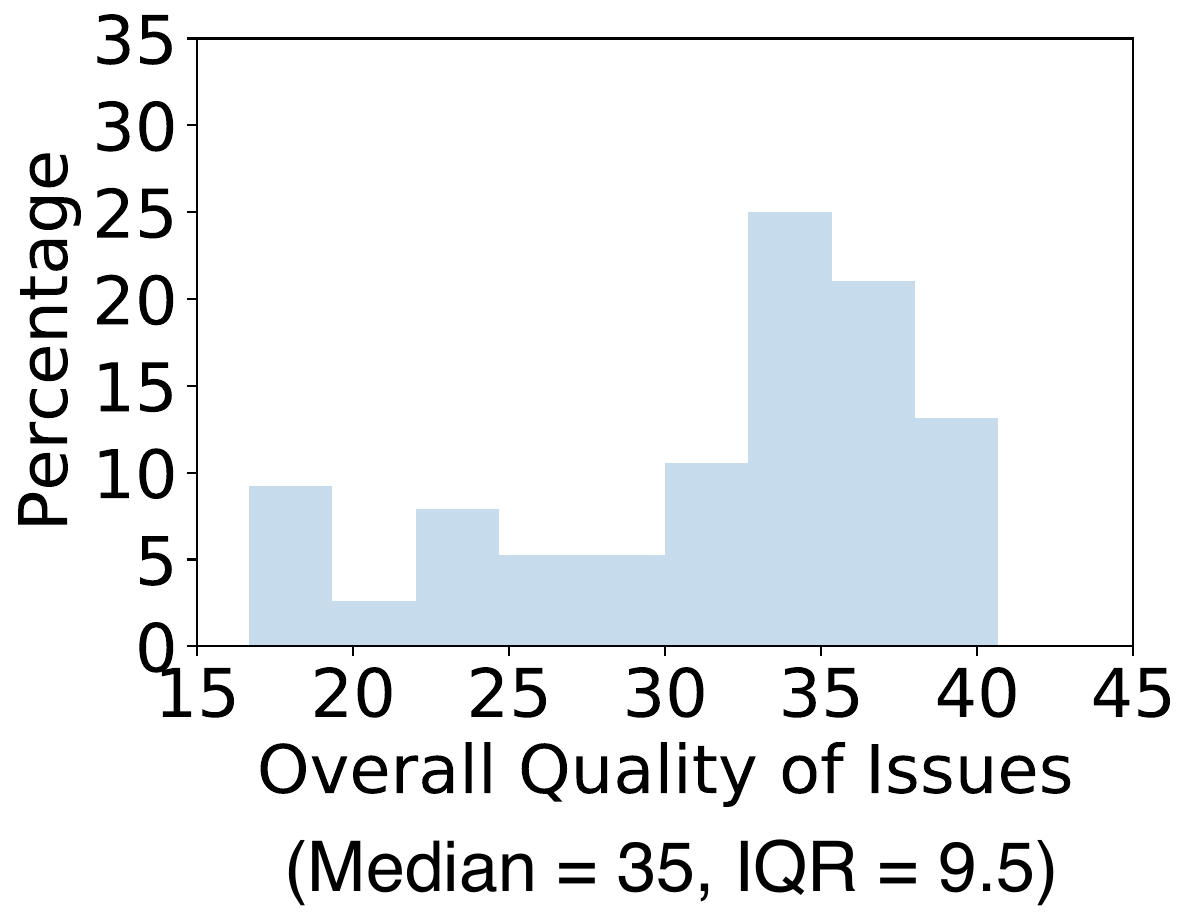}
      \caption{VSCode}
      \label{fig:overall_argument_quality_issues_vscode}
    \end{subfigure}
    \begin{subfigure}{.19\textwidth}
      \centering
      \includegraphics[width=.99\linewidth]{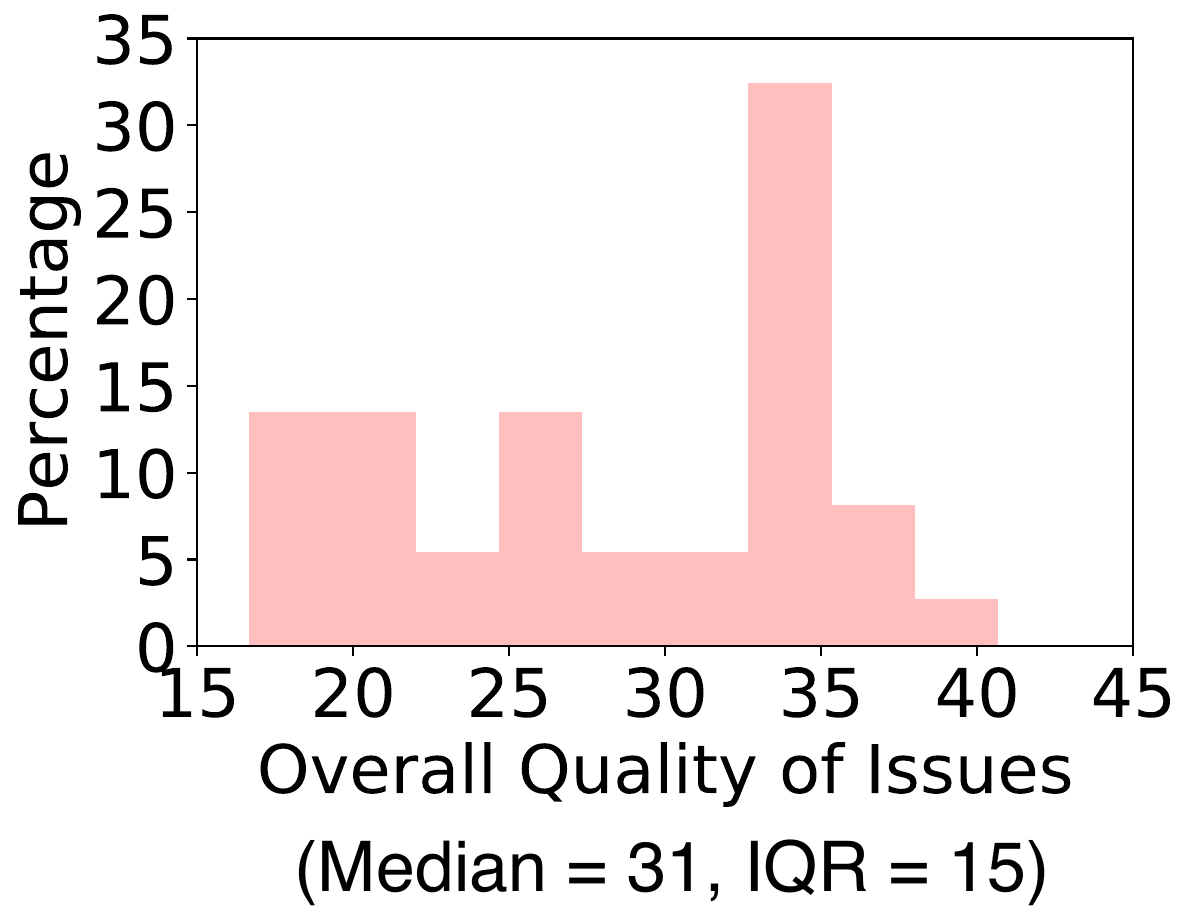}
      \caption{Atom}
      \label{fig:overall_argument_quality_issues_atom}
    \end{subfigure}
   \caption{Distribution of overall argument quality (by summing up ratings of all sub-dimensions), with the median and interquartile range (IQR), in usability issue posts for each project.}
   \Description{Five histograms are shown, each reporting the distribution of argument quality in the usability issue posts of each analyzed OSS project. The x-axes are the overall quality scores and the y-axes are the percentage distributions. The histograms for Jupyter Lab, CoCalc, and VSCode resembled a left-skewed distribution, while the histograms for Google Colab and Atom resembled a bimodal distribution.}
   \label{fig:overall_argument_quality_issues}
\end{figure*}

There are three main dimensions (\textit{cogency}, \textit{effectiveness}, and \textit{reasonableness}) in the framework~\cite{wachsmuth_computational_2017} that we used to analyze the argument quality; each contains several sub-dimensions (refer to Table~\ref{tab:argument_quality_dimensions}). 
Table~\ref{tab:argument_quality_issues} summarizes the quality of arguments that appeared in the issue posts of the five projects. For \textbf{Cogency}, we observed two tendencies: (1) arguments in issue posts of Atom, VSCode, and Google Colab more frequently demonstrated \textit{high} or \textit{low} cogency than \textit{medium} cogency, while all issues with high cogency contained premises and 96.8\% of the issues with low cogency lacked premises; and (2) in issue posts of Jupyter Lab and CoCalc, arguments generally had \textit{medium} or \textit{high} cogency. For \textbf{Effectiveness}, arguments in issue posts were generally rated as \textit{medium} in all five projects, with a small portion (roughly 10\% to 25\%) rated as \textit{high}. Looking into the sub-dimensions, \textit{Arrangement} seemed to be the most problematic, with a higher percentage of \textit{low} quality arguments, especially in Atom and Google Colab. For \textbf{Reasonableness}, again, two trends were observed: (1) in Atom and Google Colab, there was a roughly equal number of arguments that were rated as \textit{high}, \textit{medium}, and \textit{low} reasonableness, while low reasonableness was mostly contributed by a low \textit{Global Relevance}; and (2) in VSCode, CoCalc, and Jupyter Lab, the reasonableness of arguments was generally rated as \textit{medium} or \textit{high}. Figure~\ref{fig:overall_argument_quality_issues} presents the distribution of the overall quality, calculated by summing up ratings of all sub-dimensions,  across all argumentative issue posts. 

\begin{table*}[t]
  \centering
  \caption{Percentage of argument quality in comments of the longest usability issues}
  \label{tab:argument_quality-long-discussion}
  \resizebox{\textwidth}{!}{
  \begin{tabular}{@{}lccccccccccccccc@{}}
    \toprule
    & \multicolumn{3}{c}{Jupyter Lab} & \multicolumn{3}{c}{Google Colab} & \multicolumn{3}{c}{CoCalc} & \multicolumn{3}{c}{VSCode} & \multicolumn{3}{c}{Atom} \\
    \cmidrule(lr){2-4} \cmidrule(lr){5-7} \cmidrule(lr){8-10} \cmidrule(lr){11-13} \cmidrule(lr){14-16}
&High&Med.&Low&High&Med.&Low&High&Med.&Low&High&Med.&Low&High&Med.&Low \\
\hline
Local Acceptability & 30.6  & 16.1  & 53.2  & 35.3  & 17.6  & 47.1  & 48.1  & 11.1  & 40.7  & 32.2  & 17.2  & 50.6  & 37.7  & 42.6  & 19.7 \\
Local Relevance & 27.4  & 19.4  & 53.2  & 38.2  & 14.7  & 47.1  & 44.4  & 14.8  & 40.7  & 28.7  & 17.2  & 54.0  & 31.1  & 47.5  & 21.3  \\
Local Sufficiency & 24.2  & 14.5  & 61.3  & 38.2  & 14.7  & 47.1  & 44.4  & 14.8  & 40.7  & 27.6  & 11.5  & 60.9  & 27.9  & 50.8  & 21.3  \\
\rowcolor{lavender}
\textbf{Cogency}& 29.0  & 17.7  & 53.2  & 38.2  & 14.7  & 47.1  & 44.4  & 14.8  & 40.7  & 29.9  & 16.1  & 54.0  & 32.8  & 47.5  & 19.7  \\\midrule
Credibility & 45.2  & 54.8  & 0.0  & 32.4  & 67.6  & 0.0  & 37.0  & 63.0  & 0.0  & 23.0  & 77.0  & 0.0  & 41.0  & 59.0  & 0.0             \\      
Emotional Appeal & 62.9  & 37.1  & 0.0  & 38.2  & 61.8  & 0.0  & 22.2  & 77.8  & 0.0  & 37.9  & 62.1  & 0.0  & 68.9  & 31.1  & 0.0 \\
Clarity & 21.0  & 75.8  & 3.2  & 14.7  & 85.3  & 0.0  & 25.9  & 74.1  & 0.0  & 12.6  & 85.1  & 2.3  & 24.6  & 72.1  & 3.3               \\
Appropriateness & 8.1  & 83.9  & 8.1  & 11.8  & 85.3  & 2.9  & 22.2  & 74.1  & 3.7  & 2.3  & 83.9  & 13.8  & 8.2  & 91.8  & 0.0          \\
Arrangement & 27.4  & 46.8  & 25.8  & 26.5  & 47.1  & 26.5  & 18.5  & 48.1  & 33.3  & 11.5  & 47.1  & 41.4  & 29.5  & 49.2  & 21.3        \\
\rowcolor{lavender}\textbf{Effectiveness} & 21.0  & 77.4  & 1.6  & 11.8  & 88.2  & 0.0  & 11.1  & 88.9  & 0.0  & 5.7  & 93.1  & 1.1  & 23.0  & 77.0  & 0.0  \\\midrule
Global Acceptability & 25.8  & 53.2  & 21.0  & 38.2  & 38.2  & 23.5  & 37.0  & 48.1  & 14.8  & 28.7  & 36.8  & 34.5  & 32.8  & 52.5  & 14.8  \\
Global Relevance & 4.8  & 53.2  & 41.9  & 23.5  & 32.4  & 44.1  & 14.8  & 25.9  & 59.3  & 3.4  & 42.5  & 54.0  & 11.5  & 52.5  & 36.1        \\
Global Sufficiency & 14.5  & 30.6  & 54.8  & 23.5  & 32.4  & 44.1  & 29.6  & 29.6  & 40.7  & 17.2  & 18.4  & 64.4  & 18.0  & 41.0  & 41.0    \\
\rowcolor{lavender}
\textbf{Reasonableness} & 9.7  & 56.5  & 33.9  & 32.4  & 23.5  & 44.1  & 18.5  & 40.7  & 40.7  & 13.8  & 42.5  & 43.7  & 14.8  & 60.7  & 24.6  \\    \bottomrule
  \end{tabular}}
\end{table*}

\begin{figure*}[t]
    \begin{subfigure}{.19\textwidth}
      \includegraphics[width=.99\linewidth]{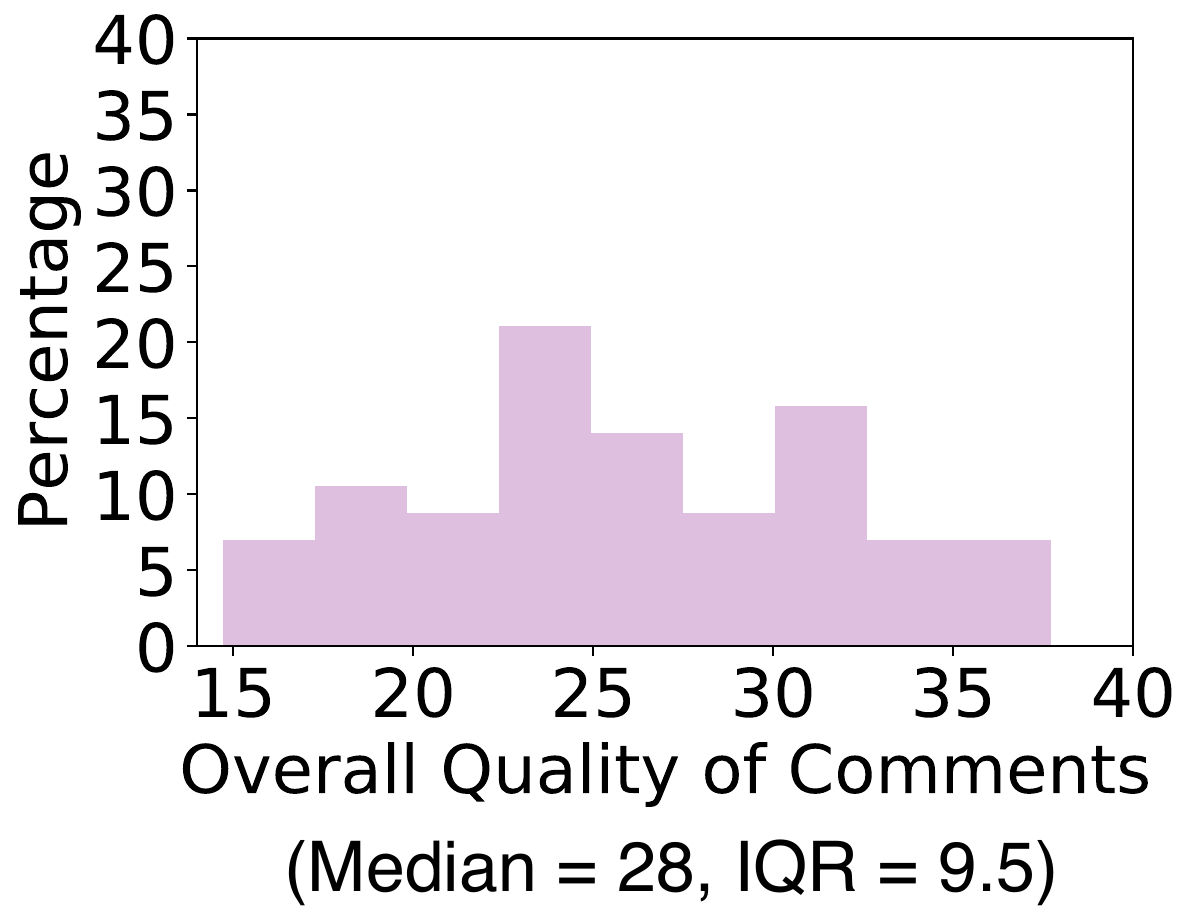}
      \caption{Jupyter Lab}
      \label{fig:overall_argument_quality_comments_jupyterlab}
    \end{subfigure}
    \begin{subfigure}{.19\textwidth}
      \centering
      \includegraphics[width=.99\linewidth]{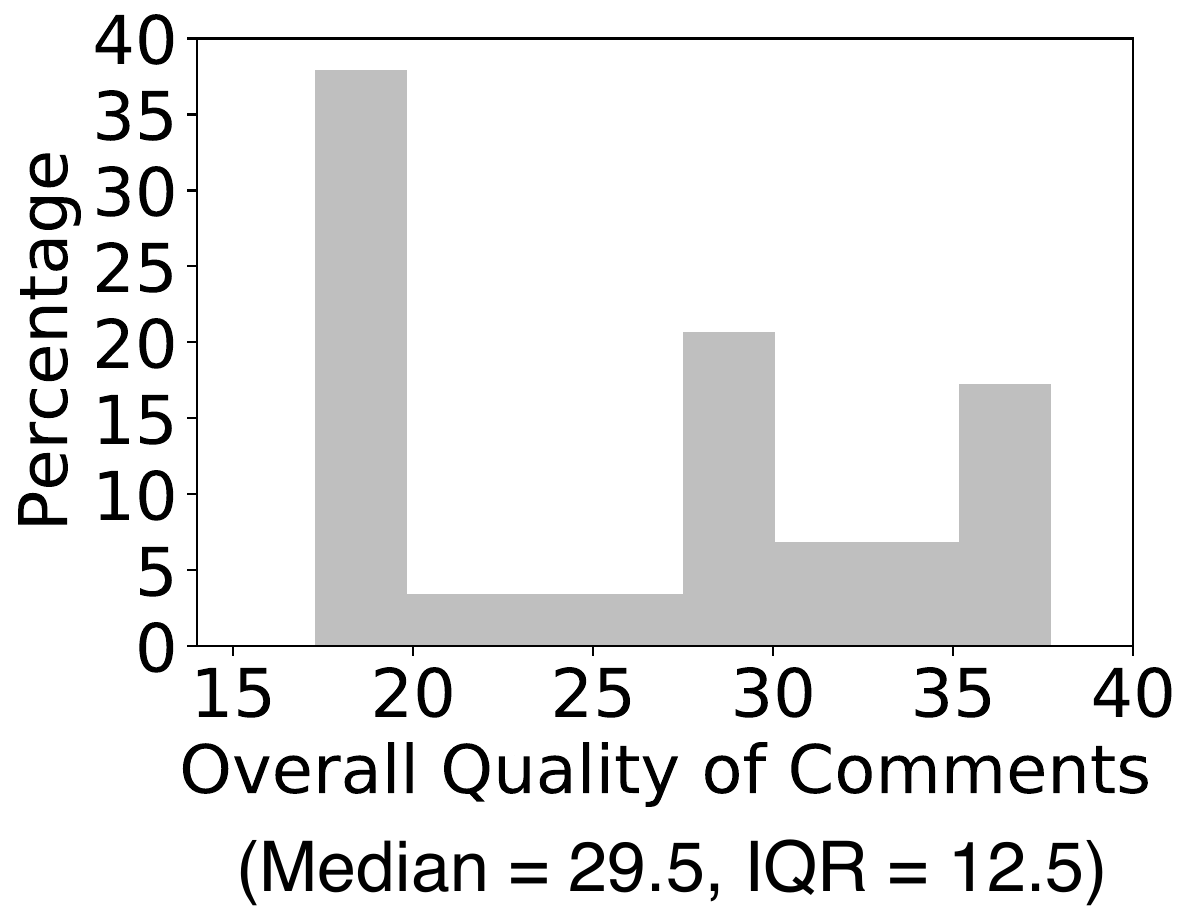}
      \caption{Google Colab}
      \label{fig:overall_argument_quality_comments_colab}
    \end{subfigure}
    \begin{subfigure}{.19\textwidth}
      \centering
      \includegraphics[width=.99\linewidth]{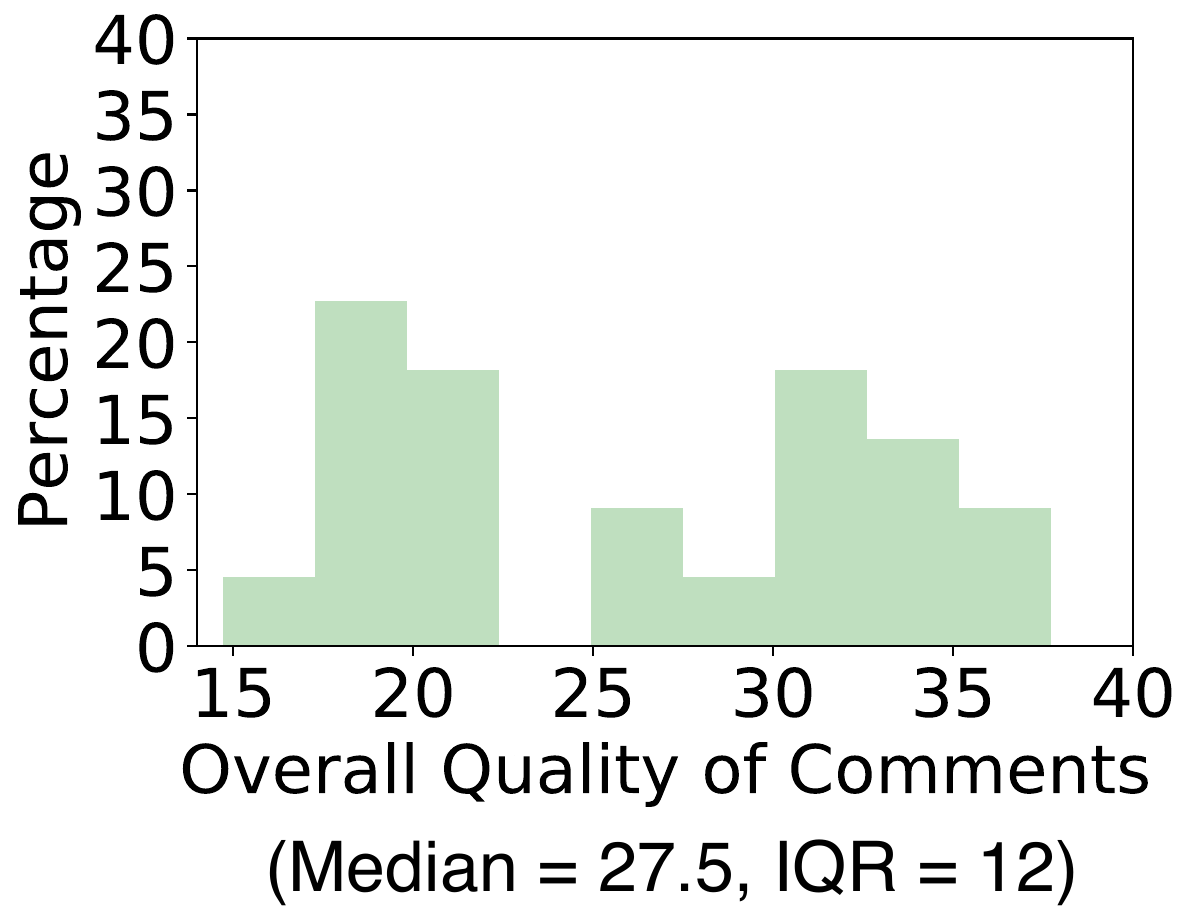}
      \caption{CoCalc}
      \label{fig:overall_argument_quality_comments_cocalc}
    \end{subfigure}
    \begin{subfigure}{.19\textwidth}
      \centering
      \includegraphics[width=.99\linewidth]{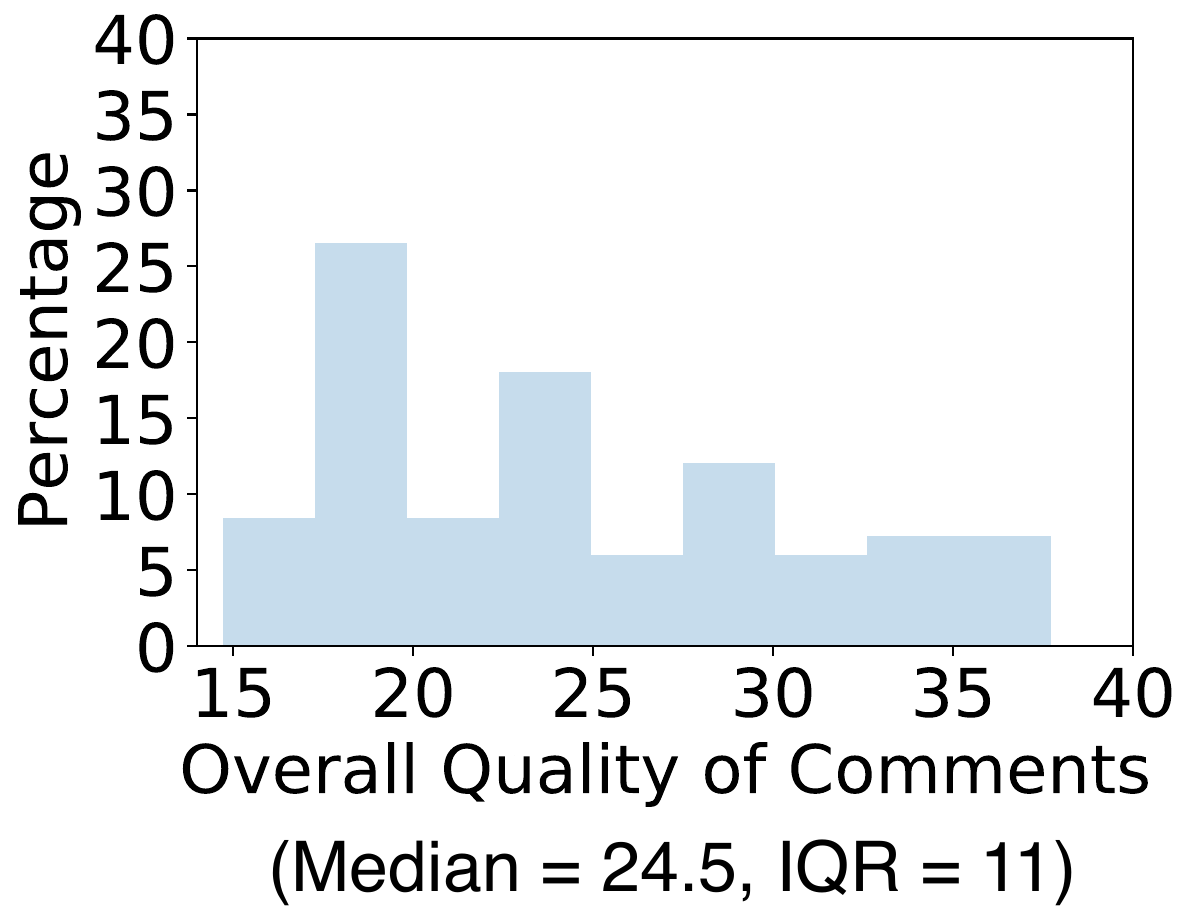}
      \caption{VSCode}
      \label{fig:overall_argument_quality_comments_vscode}
    \end{subfigure}
    \begin{subfigure}{.19\textwidth}
      \centering
      \includegraphics[width=.99\linewidth]{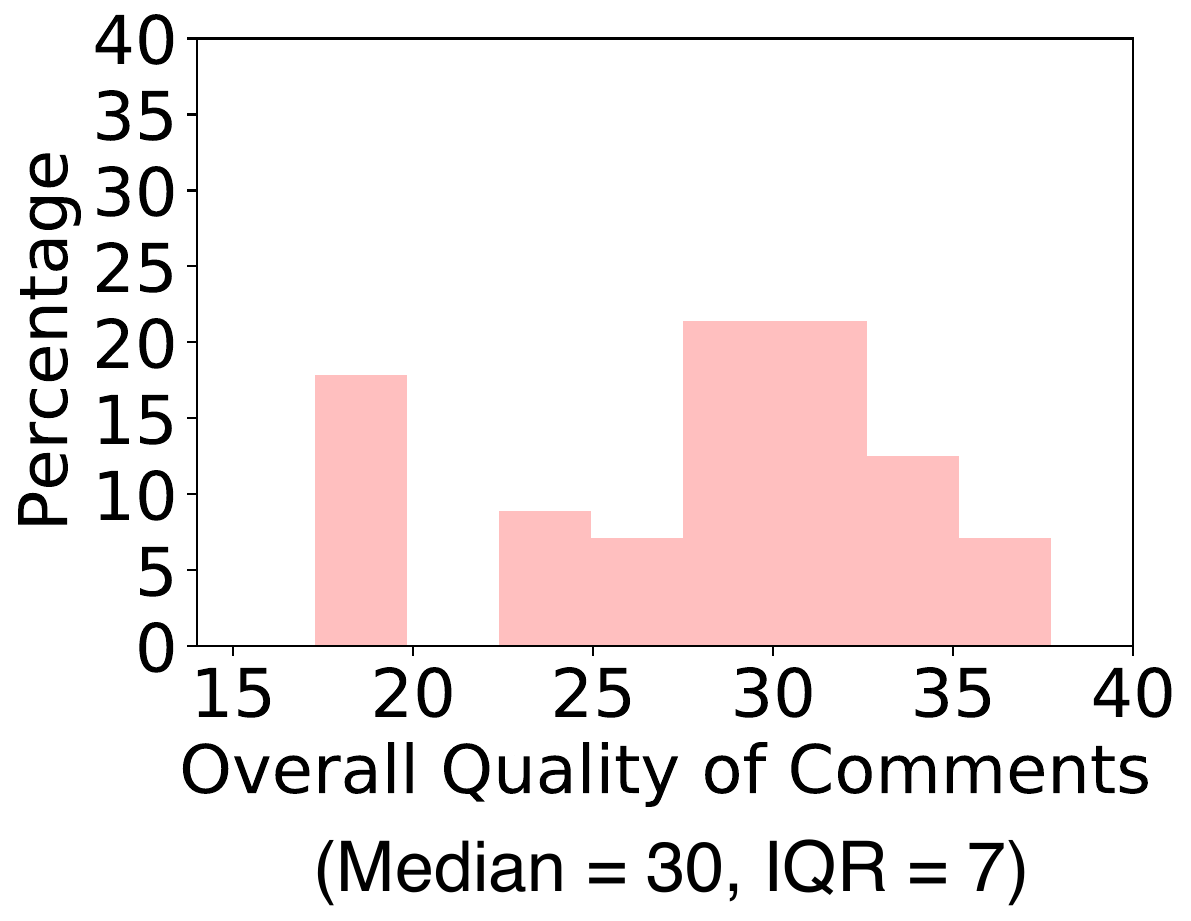}
      \caption{Atom}
      \label{fig:overall_argument_quality_comments_atom}
    \end{subfigure}
   \caption{Distribution of overall argument quality (by summing up ratings of all sub-dimensions), with median and interquartile range (IQR), in comments of the five longest usability issues for each project.}
   \Description{Five histograms are shown, each reporting the distribution of argument quality in the usability issue comments of each analyzed OSS project. The x-axes are the overall quality scores and the y-axes are the percentage distributions. The histogram resembled a normal distribution for Jupyter Lab, a bimodal distribution for Google Colab, CoCalc, and Atom, and a right-skewed distribution for VSCode.}
   \label{fig:overall_argument_quality_comments}
\end{figure*}

Table~\ref{tab:argument_quality-long-discussion} summarizes the quality of arguments in the comments of the five longest issues of each project, and Figure~\ref{fig:overall_argument_quality_comments} presents the distribution of the overall quality (by summing up ratings of all sub-dimensions) of arguments in those comments. We found that in most projects (all except Atom), the argument quality of the issue comments was generally lower than that of the issue posts. This is evident in all three quality dimensions, as well as the overall quality. For Atom, however, this trend is not observed. In this project, arguments in comments tended to have better \textit{Cogency} and \textit{Effectiveness}, but lower \textit{Reasonableness} than arguments in issue posts.

\subsection{Associations of Usability Dimensions and Argumentative Discourse With Argument Quality}
Regarding the association between usability dimensions of posted issues and argument quality, the only significant results were found on Google Colab, with the quality dimension \textit{Effectiveness} ($H = 13.86$, $p = 0.016$). This association had a considerable effect size ($\eta^2 = 0.385$). A post-hoc analysis revealed that, in Google Colab, usability issues related to \textit{\#9 Help users recognize, diagnose, and recover from errors} had significantly lower \textit{Effectiveness} than issues related to \textit{\#8: Aesthetic and minimalist design} ($p < 0.05$).

Regarding the impact of claim types, we found a statistically significant result in the VSCode project, regarding \textit{Credibility} ($H = 9.305$, $p = 0.030$), although the effect size was small ($\eta^2 = 0.089$). A post-hoc analysis revealed that usability issues with ``Hypothesis/Opinion'' claims had higher \textit{Credibility} than those with an ``Assertion'' claim ($p < 0.05$). Regarding the relationship between premise types and argument quality, we found some interesting results. Notably, statistically significant results were found on both \textit{Cogency} and \textit{Reasonableness} in most of the projects, although no significant results were found on the \textit{Effectiveness} dimension. Table ~\ref{tab: impact_premise_argument_quality} summarizes the significant results. Particularly, arguments with visual content premises tended to have higher \textit{Cogency} and \textit{Reasonableness}.

\begin{table}[t]
  \centering
  \small
      \caption{Summary of results of premise types impact on argument quality (*** p<0.001, ** p<0.01, * p<0.05)}
      \label{tab: impact_premise_argument_quality}
      \begin{tabular}{l|c|cccl}
        \toprule
        Ind. Var. & Dependent. var.  & Project & $\eta^2$  & Strength & Significant pairs\textsuperscript{\dag}  \\ \hline
        \multirow{23}{*}{\begin{sideways}Premise type\end{sideways}} & \multirow{15}{*} {Cogency} & \multirow{3}{*} {Jupyter Lab} & \multirow{3}{*} {0.38***} & \multirow{3}{*} {Large} &(Evidence < Visual)***\\
    & & & & &  (No premise < Visual)***\\
    & & & & & (Specific < Visual)***\\ 
    \cline{3-6}
    & &   \multirow{2}{*} {Google Colab} & \multirow{2}{*} {0.28***} &\multirow{2}{*} {Large}  & (No premise < Visual)* \\
    & & & & & (Specific > Visual)* \\
    \cline{3-6}
    & &   \multirow{4}{*} {CoCalc} & \multirow{4}{*} {0.52***}	& \multirow{4}{*} {Large}  &(No premise < Specific)** \\
    & & & &  &(Referring < Specific)* \\
    & & & &  &(Referring < Visual)* \\
    & & & &  &(No premise < Visual)*** \\
    \cline{3-6}
    & &   \multirow{3}{*} {VSCode} & \multirow{3}{*} {0.41***}	& \multirow{3}{*} {Large} &(Comparing < Visual)***\\
    & & & &  &(No premise < Visual)*** \\
    & & & &  &(Specific < Visual)***\\
    \cline{3-6}
    & &   \multirow{2}{*} {Atom} & \multirow{2}{*} {0.64***}& \multirow{2}{*} {Large} & (Specific < Visual)**\\
    & & & &  &(No premise < Visual)*** \\
    \cline{2-6} 
    & \multirow{8}{*} {Reasonableness} & \multirow{3}{*} {Jupyter Lab} & \multirow{3}{*} {0.37***} &  \multirow{3}{*} {Large} &(Evidence < Visual)**\\
    & & & & & (No premise < Visual)*\\
    & & & & &(Specific < Visual)***\\  
    \cline{3-6}
    &  & \multirow{2}{*} {CoCalc} & \multirow{2}{*} {0.35***}& \multirow{2}{*} {Large}  &(No premise < Visual)**\\
    & & & & &(No premise < Specific)*\\
    \cline{3-6}
    & &   \multirow{2}{*} {VSCode} & \multirow{2}{*} {0.22***} & \multirow{2}{*} {Moderate} &(No premise < Visual)**\\
    & & & & & (Specific < Visual)* \\ 
    \cline{3-6}
    & &   \multirow{1}{*} {Atom} & \multirow{1}{*} {0.572***} & \multirow{1}{*} {Large} &(No premise < Visual)*** \\
    \hline
      \end{tabular}

    \vspace{4pt}
    \dag~Promise type labels: \textit{Evidence} = Evidence that the issue is or can be resolved; \textit{Visual} = Visual content or supporting file; \textit{Specific} = Specific usage experience; \textit{Referring} = Referring to another issue.
\end{table}

 \vspace{4pt}
\begin{tcolorbox}[colframe=black,colback=gray!10,boxrule=0.5pt,arc=.3em,boxsep=-1mm]
\noindent \textbf{RQ2 Main Findings}: 
Arguments in usability issue posts tended to have high \textit{Cogency} if they included a premise, but medium \textit{Effectiveness} and \textit{Reasonableness}. Arguments in the comments of usability issues tended to have a lower quality than those in the issue posts. Furthermore, arguments supported by visual content tended to have higher \textit{Cogency} and \textit{Reasonableness}. 
\end{tcolorbox}
\section{RQ3 Results: Impact of Argument on Usability Discussion Threads}

\begin{table}[t]
    \centering
    \caption{Summary of results of discussion attributes (*** $p<0.001$, ** $p<0.01$, * p<0.05)}
    \label{tab:discussion_features}
    \resizebox{\textwidth}{!}{
    \begin{tabular}{l|llll}
        \hline
       Independen var. & Dependent var. & Project & F-value & Significant pairs  \\ \hline
       \multirow{3}{*} {Argument structure} & \multirow{1}{*} {Discussion length} & 
        \multirow{1}{*} {Atom} & \multirow{1}{*} {5.82*} & \multirow{1}{*} {(Claim \& premise > Claim Only)*} 
       \\
       & \multirow{1}{*} {\#Participants} & \multirow{1}{*} {Atom}  &   \multirow{1}{*} {4.58*} &  \multirow{1}{*} {(Claim \& Premise > Claim Only)*} \\
       & \multirow{1}{*} {\#Reactions} & \multirow{1}{*} {Jupyter Lab}&  \multirow{1}{*} {5.78*} &  \multirow{1}{*} {(Claim \& Premise < Claim Only)*}  \\
       \hline
       \multirow{5}{*} {Claim type} & \multirow{1}{*} {Discussion length}  & \multirow{1}{*} {Google Colab} &\multirow{1}{*} {5.82***} & \multirow{1}{*} {(Assertion < Probability)*}  \\ 
        & \multirow{1}{*} {\#Participants} & \multirow{1}{*} {Google Colab}  & \multirow{1}{*} {5.43**} &  \multirow{1}{*} {(Assertion < Probability)*} \\             
        & \multirow{2}{*} {Time to close} &  \multirow{2}{*} {Jupyter Lab}  &   \multirow{2}{*} {3.39*} &  \multirow{1}{*} {(Assertion < Suggestion)* } \\
        & & & & (Assertion < Hypothesis)* \\       
        & \multirow{1}{*} {\#Reactions} &   \multirow{1}{*} {Google Colab} & \multirow{1}{*} {7.64***} &  \multirow{1}{*} {(Assertion < Probability)**}                \\
        \hline
       \multirow{3}{*} {Premise type} & \multirow{1}{*} {Time to first comment}  & \multirow{1}{*} {Google Colab} &\multirow{1}{*} {3.29*} & \multirow{1}{*} {(No premise < Specific)*}  \\ 
        & \multirow{2}{*} {Time to close} & \multirow{2}{*} {VSCode}  & \multirow{2}{*} {5.49***} &  \multirow{1}{*} {(No premise > Clarification)***} \\ 
        & & &  &  \multirow{1}{*}  {(Visual < Clarification)***} \\
        \hline
        \multirow{1}{*} {Quality: Reasonableness} & \multirow{1}{*} {Time to close}  &  \multirow{1}{*} {Jupyter Lab} &  \multirow{1}{*} {4.89*} &  \multirow{1}{*} {(Medium > High)* } \\
        \multirow{2}{*} {Quality: Effectiveness} & \multirow{2}{*} {\#Reactions} 
 & \multirow{1}{*} {VSCode} & \multirow{1}{*} {6.61**} &  \multirow{1}{*} {(Medium < High)**}  \\
       &  &  \multirow{1}{*} {Atom} & \multirow{1}{*} {9.13***} &  \multirow{1}{*} {(Medium < High)**} \\
        \hline
   \end{tabular}}
\end{table}

To answer RQ3, we examined the impact of the argument discourse and quality dimensions on several attributes of usability issue discussions. The summary of the results is presented in Table~\ref{tab:discussion_features}. 

\textit{Argument structure:} Several significant results were found for this independent variable. Particularly, in the Atom project, arguments with the claim and premise in the issue posts were associated with longer discussions, involving more participants, when compared to arguments with only the claim. In the Jupyter Lab project, issue posts that contained arguments with the claim and premise received a lower number of reactions than those with only the claim. So, it seemed that the inclusion of a premise in the argument encouraged participants to voice their opinions in the comments, instead of using simple reaction emojis.
    
\textit{Claim type:} Upon analyzing the association between the type of claim and discussion attributes, the significant results appear in Google Colab and Jupyter Lab. Preliminary evidence suggests that compared to the \textit{Probability/doubtfulness} claim type, arguments with the \textit{Assertion} claim type were not only associated with shorter discussion and less participation of collaborators but also received fewer reactions from other contributors. At the same time, in Jupyter Lab, arguments with the \textit{Assertion} claim type were associated with a shorter time for an issue to get closed, compared to \textit{Hypothesis/opinion} and \textit{Suggestion/recommendation} claim types. So, it seemed that issues posted with assertion claims would involve less community input but would be resolved faster.
    
\textit{Premise type:} For this independent variable, significant values were observed in the Google Colab and VSCode projects. In Google Colab, issues posted without a premise got their first comments faster; one possible reason was that the community might have asked those issues to provide more information. At the same time, in VSCode, issues posted with the premise type of \textit{Visual content or supporting file} were resolved the fastest, confirming the value of using visual content in usability issues.
    
\textit{Argument quality dimensions:} There were two out of the three main argument quality dimensions that received significant results. Particularly, issue posts that contained arguments with high \textit{Reasonableness} in Jupyter Lab led to a faster resolution; this is expected since high reasonableness indicates that the argument is more likely to be accepted and provides valuable solutions. Besides, arguments with high \textit{Effectiveness} in VSCode and Atom were associated with more reactions to the issue posts; many of those reactions were thumbs-up and plus-one emojis, indicating an agreement to the argument.

\vspace{4pt}
\begin{tcolorbox}[colframe=black,colback=gray!10,boxrule=0.5pt,arc=.3em,boxsep=-1mm]
\noindent \textbf{RQ3 Main Findings}: Significant results were found in several projects, indicating that (1) including a premise in the argument encouraged commenting and discouraged the use of simple reaction emojis, (2) using an assertion claim was associated with less community input but faster resolution, and (3) arguments with higher reasonableness and higher effectiveness were more acceptable by the community.
\end{tcolorbox}

\section{Discussion}
In this study, we focused on compiling an initial profile of argumentative usability issue discussions in OSS. In the five OSS projects that we investigated, we found that usability discussion participants widely employed arguments (i.e., an average of 93.7\% in issue posts and 69.3\% in comments). Our results provide several valuable insights into the role of argument discourse and quality in this collaborative endeavor made by OSS community members. These insights can also inform other types of distributed, asynchronous collaborative work that involves diverse participants.

\subsection{Addressing the Shortage of Collective Intelligence About OSS Usability.}
Our results indicated that, compared to the comments, usability issue posts did not only contain more arguments, but those arguments were also more likely to be supported by a premise. Arguments in issue posts also generally had better quality than those in the comments. Moreover, while both issue posters and commenters frequently expressed confidence with \textit{assertion} claims, issue posters more frequently used \textit{suggestion/recommendation} claims, which were often firmer than the commenters' \textit{hypothesis/opinion} claims. When compared to those in the issue posts, premises in comments also used less \textit{specific usage experiences} and less \textit{visual content or supporting files}, two elements that are strongly associated with higher argument quality. 

Although it is commendable that usability issue posters frequently used effective argumentation to clarify their points of view, the lower community engagement in using high-quality argumentative devices in comments to explore different perspectives is a potentially concerning situation. This is particularly because usability arguments do not only support the decision-making for developers, but they also serve as a source of collective intelligence~\cite{Suran2020} to educate and inform OSS developers about the diverse needs of their users. The benefits of involving diverse points of view to create collective intelligence are highlighted in other collaborative platforms such as community-based question answering~\cite{Liu2023} and social media~\cite{Baughan2021}. However, this problem for OSS usability is unique and important because the OSS communities often lack awareness of usability concerns~\cite{Wang2022IEEESoftware, nichols2006usability}, which can be very well provided by the currently insufficient collective argumentative exchanges. Although arguments were frequently used in usability discussions, their qualities varied. Thus, we still need to support a wider range of OSS community members to engage in high-quality argumentative discourse around usability to raise awareness and drive improvements. Future studies should focus on this aspect in OSS communities and in other online collaboration contexts that have imbalanced power distributions among community members and/or need to include the participation of marginalized users.

\subsection{Fulfilling the Prominent Need of Visual Communication in Usability Arguments}
One prominent finding from this research is related to the visual content used in usability arguments. The majority of participants in issue discussion threads supported their stance with the help of \textit{visual content or supporting files}. We also found a strong correlation between the use of this type of premise and the argument quality dimensions. Specifically, our study revealed that using visual content or external files tended to result in high-quality arguments that has high \textit{cogency} and \textit{reasonableness}. Echoing the findings from previous work~\cite{Agrawal_2022visual, sanei2023characterizing}, these results highlighted the importance of visual communication in argumentative usability discussions.

While it may seem obvious, the popularity and the positive effects of using visual content in usability discussions are non-trivial. This is because GitHub Issues is a primarily text-based platform. The features for supporting the use of visual content are very limited. Our results did not only reveal the prominent need for using visual content in making usability arguments but we also provided concrete empirical evidence on the relationship between the usage of these materials and the argument quality. These results indicated that OSS issue-tracking systems should include more visual communication support to facilitate effective usability discussions. At the same time, tools and techniques should be explored to better integrate visual communication into the argumentation discourse to directly support the stakeholders' collective effort in mutual understanding and negotiation to improve OSS usability.

\subsection{Tackling the Complex Factors Influencing Usability Argument Quality.}
Our approach of adapting Wachsmuth et al.'s framework~\cite{wachsmuth_computational_2017} to evaluate argument quality provided a comprehensive view of factors that contribute to the discussion quality. For example, we identified that low cogency (i.e., claims that are weakly supported) in usability-related arguments was largely due to a simple lack of premises. Similarly, poor arrangement was often the factor that hurt the argument's effectiveness (i.e., the power to persuade others), and low global relevance (i.e., inability to provide useful insights) was frequently associated with low reasonableness. When considering ways to support OSS community members to improve their arguments in usability discussions, these insights allow us to hone in our focus to address the most important factors (e.g., lack of premises or poorly arranged discourse). These results can also be applied beyond OSS usability to future investigations aimed at helping users of other online collaboration platforms (e.g., team messaging and collaboration tools like Slack and textual/visual collaboration platforms like Miro) make more effective and impactful contributions.

Interestingly, while previous work highlighted the importance of using templates to create issues and bug reports in OSS~\cite{LI2023}, we frequently encountered low-quality arguments related to usability that used a template as a shortcut but did not provide coherent information. Similarly, while empathy towards users is often emphasized as an important factor to effectively address user needs in software design and development~\cite{Wright2008, Gunatilake2024}, we found that overly concentrating on personal experiences is often not a very compelling way to make strong arguments. These observations highlighted the difficulty for non-developers to provide convincing and high-quality usability arguments. Contributing to the findings by \citet{Hellman2022Characterizing} and \citet{Sanei2025}, which emphasized the importance of improving the engagement of end-users and designers in OSS, our results highlighted the need to enhance the their ability to make effective and meaningful arguments to achieve true engagement.

Moreover, previous research has indicated that community characteristics such as size~\cite{Hwang2021}, diversity~\cite{Dubois2022}, and communication norms~\cite{Dym2020} may also have impact on discussion quality, including argument quality. Although our study did not directly analyze these factors, we found that usability issue posts in Google Colab, VSCode, and Atom tended to less frequently include a premise (see Figure~\ref{fig:Argumentative-Type-Issue}) and as a result, had more arguments with low cogency (see Table~\ref{tab:argument_quality_issues}). One explanation is that these three projects are all managed by big corporations (Google, Microsoft, and GitHub, respectively). Thus, usability-related discussions may be handled internally and the issue tracking system is used to manage decisions coming out of those internal discussions as tasks. As we previously identified and echoing prior work on commercial participation in OSS~\cite{Osborne2025}, this centralized approach may miss the opportunity to harness shared knowledge and insights from the broader community.

\subsection{Limitations and Future Work}
Our study has several limitations that can be addressed in future work. First, we acknowledge that this research only analyzed a limited number of usability issues, relying on an existing dataset~\cite{sanei2023characterizing}. For argument quality analysis in the issue comments, we also only focused on the five longest issues in each project to manage manual effort while capturing discussion complexity. To enhance generalizability, future research should aim to expand the sample of usability issues. A larger dataset might also cover a wider range of usability dimensions, including those that are rarely touched on in our dataset. Second, this study only investigated popular OSS projects. The usability arguments in smaller sizes of OSS projects may have different characteristics and should be addressed in future work. Similarly, we did not explicitly analyze how project-specific factors (e.g., project size, project maturity, community norm, etc.) and contributor's profile (e.g., background and expertise) may contribute to the difference in argument dynamics and qualities. This can also be a focus in future work. Finally, we mainly focused on analyzing argumentative discourse and quality based on the artifacts created by OSS community members. While the information and insights gathered through this approach were rich, we were not able to consider personal aspects such as motivations, background, native language, and writing style of the OSS community members. To triangulate the findings of this study, future research could involve user studies (e.g., interviews and surveys) and experiments (e.g., studies analyzing participants' cognitive process of writing arguments or reactions to different argument discourses and qualities) with contributors to OSS usability discussions.

\section{Conclusion}
The overall objective of this study was to characterize the argumentative structure and quality dimensions of OSS usability issues and identify their correlations with other collaboration attributes. To achieve this, a comprehensive approach combining qualitative and quantitative methods was employed to examine usability discussions within five OSS projects. We found that OSS contributors extensively infused their usability discussions with argument discourses, although their quality varied. Moreover, arguments in issue posts tended to have a higher level of quality than those in the comments. OSS community members also frequently used visual content to support their claims, which resulted in higher-quality arguments. Additionally, high-quality arguments, particularly those in the issue posts, had a positive impact, influencing the discussion participants' overall behavior. Our results provided implications to address the shortage of collective intelligence about usability in the OSS communities. We pointed out possible ways to help OSS community members make more effective usability arguments while highlighting the complexity of this issue. These insights can support future research aimed at fostering diverse members to participate in distributed and asynchronous collaborative communities.


\begin{acks}
This work is partially supported by the Alfred P. Sloan Foundation (G-2021-16745) and the Natural Sciences and Engineering Research Council of Canada (RGPIN-2018-04470).
\end{acks}




\bibliographystyle{ACM-Reference-Format}
\bibliography{references}


\end{document}